\documentclass[useAMS,usenatbib]{mn2e}

\usepackage{graphicx}

\newcommand{\BH}{\ensuremath{\mathrm{BH}}}
\newcommand{\MBH}{\ensuremath{M_\mathrm{BH}}}
\newcommand{\lMBH}{\ensuremath{\log \MBH}}

\newcommand{\sigMBH}{\ensuremath{\Delta(\MBH)}}
\newcommand{\sigBH}{\sigMBH}
\newcommand{\rhoBH}{\ensuremath{\rho_\mathrm{BH}}}
\newcommand{\rhoBHunits}{\ensuremath{\,\times 10^5\Msun\MPC\3}}
\newcommand{\BHMF}{\ensuremath{\mathrm{BHMF}}}
\newcommand{\eff}{\ensuremath{\epsilon}}

\newcommand{\Msun}{\ensuremath{~\mathrm{M}_\odot}}
\newcommand{\Lsun}{\ensuremath{~\mathrm{L}_\odot}}

\newcommand{\LEDD}{\ensuremath{L_\mathrm{Edd}}}

\newcommand{\LKsun}{\ensuremath{~\mathrm{L}_{K\odot}}}

\newcommand{\NBH}{\ensuremath{N(M,t)}}
\newcommand{\NAGN}{\ensuremath{\phi(L,t)}}
\newcommand{\logL}{\ensuremath{\log L}}
\newcommand{\dlogL}{\ensuremath{\mathrm{d}\log L}}
\newcommand{\dM}{\ensuremath{\mathrm{d}M}}
\newcommand{\Mdot}{\ensuremath{\dot M}}
\newcommand{\MdotBH}{\ensuremath{\Mdot(M,t)}}

\newcommand{\zo}{\ensuremath{z_s}}
\newcommand{\tzo}{\ensuremath{t_s}}

\newcommand{\Msph}{\ensuremath{M_\mathrm{bul}}}
\newcommand{\Lsph}{\ensuremath{L_\mathrm{bul}}}
\newcommand{\Lbul}{\ensuremath{\Lsph}}

\newcommand{\LKsph}{\ensuremath{L_\mathrm{K,bul}}}
\newcommand{\LbulK}{\LKsph}

\newcommand{\sigstar}{\ensuremath{\sigma_\star}}

\newcommand{\MBHsig}{\MBH-\sigstar}
\newcommand{\MBHlum}{\MBH-\Lbul}

\newcommand{\rs}{\ensuremath{r^\star}}
\newcommand{\zs}{\ensuremath{z^\star}}
\newcommand{\ten}[1]{\ensuremath{10^{#1}}}
\newcommand{\xten}[1]{\ensuremath{\times 10^{#1}}}
\newcommand{\1}{\ensuremath{^{-1}}}
\newcommand{\2}{\ensuremath{^{-2}}}
\newcommand{\3}{\ensuremath{^{-3}}}
\newcommand{\CM}{\ensuremath{\mathrm{\,cm}}}
\newcommand{\MIC}{\ensuremath{\,\mu\mathrm{m}}}
\newcommand{\KM}{\ensuremath{\mathrm{\,km}}}

\newcommand{\MPC}{\ensuremath{\mathrm{\,Mpc}}}
\newcommand{\SEC}{\ensuremath{\mathrm{\,s}}}
\newcommand{\YR}{\ensuremath{\mathrm{\,yr}}}

\newcommand{\ERG}{\ensuremath{\mathrm{\,erg}}}

\newcommand{\KEV}{\ensuremath{\mathrm{\,keV}}}

\title[Local Black Holes, AGN Relics and the XRB]{Local Supermassive Black
Holes, Relics of Active Galactic Nuclei and the X-ray Background}
\author[A. Marconi et al.]{
A. Marconi$^{1}$\thanks{E-mail:marconi@arcetri.astro.it},
G. Risaliti$^{1,2}$,
R. Gilli$^{1}$,
L. K. Hunt$^{3}$,
R. Maiolino$^{1}$,
and M. Salvati$^{1}$ 
\\
$^{1}$INAF-Osservatorio Astrofisico di Arcetri,
Largo Fermi 5, I-50125 Firenze, Italy\\
$^{2}$Harvard-Smithsonian Center for Astrophysics,
60 Garden street, Cambridge, MA 02138, USA\\
$^{3}$INAF-Istituto di Radioastronomia-Sez.~Firenze,
	      Largo Fermi 5, I-50125 Firenze, Italy}
\begin{document}

\date{Accepted ... . Received ... ; in original form ... }

\pagerange{\pageref{firstpage}--\pageref{lastpage}} \pubyear{2003}

\maketitle

\label{firstpage}

\begin{abstract}
We quantify the importance of mass accretion during AGN phases in the growth of
supermassive black holes (\BH) by comparing the mass function of black holes in
the local universe with that expected from AGN relics, which are black holes
grown entirely with mass accretion during AGN phases.
The local \BH\ mass function (\BHMF) is estimated by applying the well-known
correlations between \BH\ mass, bulge luminosity and stellar velocity
dispersion to galaxy luminosity and velocity functions.  We find that different
correlations provide the same \BHMF\ only if they have the same intrinsic
dispersion. The density of supermassive black holes in the local universe which
we estimate is $\rhoBH = 4.6_{-1.4}^{+1.9}\, h_{0.7}^2 \rhoBHunits$.
The relic \BHMF\ is derived from the continuity equation with the only assumption
that AGN activity is due to accretion onto massive \BH's and that merging is
not important.  We find that the relic \BHMF\ at $z=0$ is generated mainly at
$z<3$ where the major part of \BH's growth takes place. Moreover, the \BH\ 
growth is anti-hierarchical in the sense that smaller \BH's
($M_{BH}<10^{7}M_\odot$) grow at lower redshifts ($z<1$) with respect to more
massive one's ($z\sim 1-3$).
Unlike previous work, we find that the \BHMF\ of AGN relics is perfectly
consistent with the local \BHMF\ indicating the local black holes were 
mainly grown
during AGN activity.  This agreement is obtained while satisfying, at the same
time, the constraints imposed from the X-ray background.  The comparison
between the local and relic \BHMF's also suggests that the merging process is
not important in shaping the relic \BHMF, at least at low redshifts ($z<3$),
and allows us to estimate the average radiative efficiency ($\varepsilon$), the
ratio between emitted and Eddington luminosity ($\lambda$) and the average
lifetime of active \BH's.
Our analysis thus suggests the following scenario: local black holes grew
during AGN phases in which accreting matter was converted into radiation with
efficiencies $\varepsilon = 0.04-0.16$ and emitted at a fraction $\lambda =
0.1-1.7$ of the Eddington luminosity. The average total lifetime of these
active phases ranges from $\simeq 4.5\times 10^{8}$ yr for
$M_{BH}<10^{8}M_\odot$ to $\simeq 1.5\xten{8}$ yr for
$M_{BH}>10^{9}M_\odot$ but can become as large as $\sim \ten{9}\YR$
for the lowest acceptable $\eff$ and $\lambda$ values.
\end{abstract}


\section{Introduction}

Since the discovery of quasars (\citealt{schmidt}) 
it has been suggested that Active
Galactic Nuclei (AGN) are powered by mass accretion onto a supermassive Black
Hole (\BH) with mass in the range \ten{6}-\ten{10}\Msun\ (\citealt{salpeter,
zeldovich, lyndenbell}).  This paradigm combined with the observed
evolution of AGNs implies that a significant number of galaxies in the local
universe should host a supermassive \BH\ (e.g.~\citealt{soltan, cavaliere89,
chokshi}).

Supermassive
\BH's are now detected in $\sim 40$ galaxies through gas and stellar dynamical
methods (\citealt{kormendy, merritt}).  Some galaxies
are quiescent (e.g.~M32, \citealt{vandermarel};
NGC 3250, \citealt{barth})
and some are mildly or strongly active (e.g.~M87, \citealt{marconi97,macchetto}; Centaurus A, \citealt{marconi01};  Cygnus A, \citealt{tadhunter}).
The mass of the \BH\ correlates with some properties of the host
galaxy such as spheroid (bulge) luminosity and mass (\citealt{kr95,magorrian}),
light concentration (Graham et al. 2001) 
and with the central stellar velocity dispersion
(\citealt{fm00, gebhardt}).  The latter correlation was
thought to be the tightest but \cite{MH03} have recently shown that
all the correlations have similar intrinsic dispersion 
when considering only galaxies with secure BH detections
(see also \citealt{mclure, erwin}).
Overall, the dispersion is of the order of 0.3 in \lMBH\
at a given value of \Lsph, \Msph\ or \sigstar. 
The existence of any correlations of \BH\ and host galaxy bulge properties has
important implications for theories of galaxy formation in general and bulge
formation in particular. Indeed, several attempts at explaining the origin of
these correlations and the difficulties/constraints that they pose to galaxy
formation models can be found in the literature
(e.g.~\citealt{silk98,cattaneo99,haehnelt00,ciotti01,cavaliere02,adams03},
and references therein).

To date, all newly found \BH's
have masses (or upper limits) in agreement with those expected from the above
correlations suggesting that all galaxies host a massive \BH\ in their nuclei.
By applying the correlations between \MBH\ and host galaxy properties it is
then possible to estimate the mass function of local \BH's or, more simply,
their total mass density (\rhoBH) in the local universe (e.g.~\citealt{salucci,MS02, yu02, ferrarese, aller}).

It is important to verify if local BH's are exclusively relics of AGN
activity or if other mechanisms, such as merging,
play a role. Hereafter we will call 'AGN relics', or simply
relics, those black holes which grew up from small seeds
(1-\ten{3}\Msun) following mass accretion during AGN phases.
For instance, an AGN relic of \ten{9}\Msun\ is different from
a \BH\ of the same mass but which was formed by the merging of many
smaller \BH's.

A simple comparison between local and relic BH's was performed by
\cite{salucci} and \cite{fi99} who determined \rhoBH\ from the observed X-ray
background emission.  A revised estimate was obtained by \cite{erz02} who found
$\rhoBH =  (7.5 - 16.8)\xten{5}\, (\eff/0.1)^{-1}\rhoBHunits$ in disagreement
with the estimate from local black holes ($\rhoBH=(2.5\pm 0.4)\,
h^2_{65}\rhoBHunits$, \citealt{yu02,aller}).  \cite{erz02} thus suggested that,
in order to reconcile this discrepancy, massive \BH's should have large
accretion efficiencies (i.e.~larger than the canonically adopted value of $\eff
=0.1$), hence they should be rapidly rotating.  A more detailed comparison was
performed by \cite{MS02} who found an agreement between the black hole mass
functions (hereafter \BHMF) of local and relic black holes.  Recently, however,
\cite{yu02} and \cite{ferrarese} found a disagreement at large masses ($\MBH\ge
\ten{8}\Msun$) where more AGN relics are expected relative to local \BH's. As
previously stated, a possibility to reconcile this discrepancy is to assume
high accretion efficiencies but, clearly, this issue is still much debated.

The relation between AGN relics and local \BH's is also being studied in the
framework of coeval evolution of \BH\ and host galaxy.  Several physical models
have been proposed in which the fueling of the \BH, hence the AGN activity, is
triggered by merging events (in the context of the hierarchical structure
formation paradigm, see for instance 
\citealt{kauffmann00,volonteri03,menci03,wyithe03,hatziminaoglou03,haehnelt03})
or is simply directly related to the star formation history of the host galaxy
(e.g.~\citealt{dimatteo03,granato04,haiman03}).  The \BH\ has then a feedback
on the host galaxy through the energy released in the AGN phase
(e.g.~\citealt{silk98,blandford99,begelman}).  As a result of this double interaction
(galaxy feeding the BH -- AGN feedback on the galaxy), these models can in
general reproduce both the observed BH-host galaxy correlations and the AGN
luminosity functions (e.g. \citealt{haehnelt98,monaco00,nulsen00} and previous
references).  However, this big effort in modeling cannot uniquely answer the
question if local \BH's are relics of AGN's, since a wide range of models with
many different underlying assumptions cannot be ruled out with the
available observational constraints.
 
The aim of this paper is to investigate the assumption that massive black holes in
nearby galaxies are relics of AGN activity by comparing the local \BHMF\ with
that of AGN relics.  We remark that in this paper we do not build a
physical model of the coevolution between central \BH\ and host galaxy
but we compare differential and integrated mass densities (local \BH's)
with differential and integrated energy densities (AGN's),
with the only assumption that AGN activity is caused by mass accretion onto
the central \BH.

We refine the analysis by \cite{MS02} and evaluate
the discrepancies found by other authors between local and relic \BHMF's.  In
Section \ref{sec:lbhmf} we estimate the local \BHMF\ by applying the known
correlations between \MBH\ and host galaxy properties to the galaxy luminosity
and stellar velocity dispersion function.  We also check the self-consistency
of the results, and show that different \MBH--host-galaxy-properties relations provide
the same \BHMF\ within the uncertainties.  In Section \ref{sec:rbhmf} we use
the continuity equation to estimate the \BHMF\ of AGN relics and in Section
\ref{sec:localrelics}, we compare local and relic \BHMF's, and find that local
BH's are consistent with AGN relics.  We then show (Sec. \ref{sec:xrb}) that
the energetic constraints inferred from the X-ray background (XRB)
are also satisfied
and that there is no discrepancy between \rhoBH\ of local BH's and that
expected from the XRB. In Sections \ref{sec:efflambda}, \ref{sec:hist}, and
\ref{sec:time} we discuss constraints on the accretion efficiency (\eff) and on
the Eddington ratio ($\lambda= L/\LEDD$, where $L$ is the AGN luminosity and
\LEDD\ is the Eddington luminosity of the active BH), and we estimate the
accretion history and the average lifetime of massive BH's.  We summarize our
results and we draw our conclusions in Section \ref{sec:conclusions}.

In this paper we adopt the current ``standard'' cosmological model, with
$H_0=100 h\,\KM\SEC\1\MPC\1$ and $h=0.7$,
$\Omega_M=0.3$ and $\Omega_\Lambda=0.7$.

\begin{table*}
 \centering
 \begin{minipage}{0.75\linewidth}
  \caption{\label{tab:galtypes}Adopted galaxy properties
  per morphological types.}
  \begin{tabular}{@{}lllllr@{}}
  \hline
  			& E & S0 & Sab & Scd & Notes \\
  \hline
  Morph. Fraction	& 0.11$\pm$0.03 & 0.21$\pm$0.05 & 0.43$\pm$0.07 & 0.19$\pm$0.07	& 1 \\
  m(Bulge)-m(Total)	& 0		& $0.64\pm0.30$ & $1.46\pm0.56$ & $2.86\pm0.59$ & 2 \\
  m(Bulge)-m(Total)	& 0		& $0.60\pm0.50$ & $1.78\pm1.01$ & $2.78\pm1.21$ & 3 \\
  \hline
 \end{tabular}\\
  1. From \protect\citealt{fukugita98} (see text for details).\\
  2. B band. Estimate by \protect\cite{aller} based on data from
  \protect\citealt{simien}.\\
  3. H band. Based on data from \citealt{hunt04}.\\
\end{minipage}
\end{table*}

\section{The Mass Function of Local Black Holes}
\label{sec:lbhmf}

The first step of the analysis presented here consists in 
the determination of the mass function of local BH's,
i.e.~black holes residing in nearby galaxies. 
The sample of galaxies with dynamically measured \BH\ masses is small ($\sim
40$) and not selected with well defined criteria. Thus it is useless
for a direct determination of the local black hole mass function. However,
the \BHMF\ can be derived by applying the
existing relations between \MBH\ and host galaxy properties 
to galaxy luminosity or velocity functions.
After we have described the adopted formalism, we will verify the
consistency of the \MBHsig\ and \MBHlum\ relations in providing the
same \BHMF\ within the uncertainties. We will then estimate the
\BHMF\ for early and all galaxy types, showing that 
different galaxy luminosity/velocity functions provide \BHMF's
which are in agreement within the uncertainties.

\subsection{Formalism}
\begin{figure*}
\centering
\includegraphics[width=0.48\linewidth]{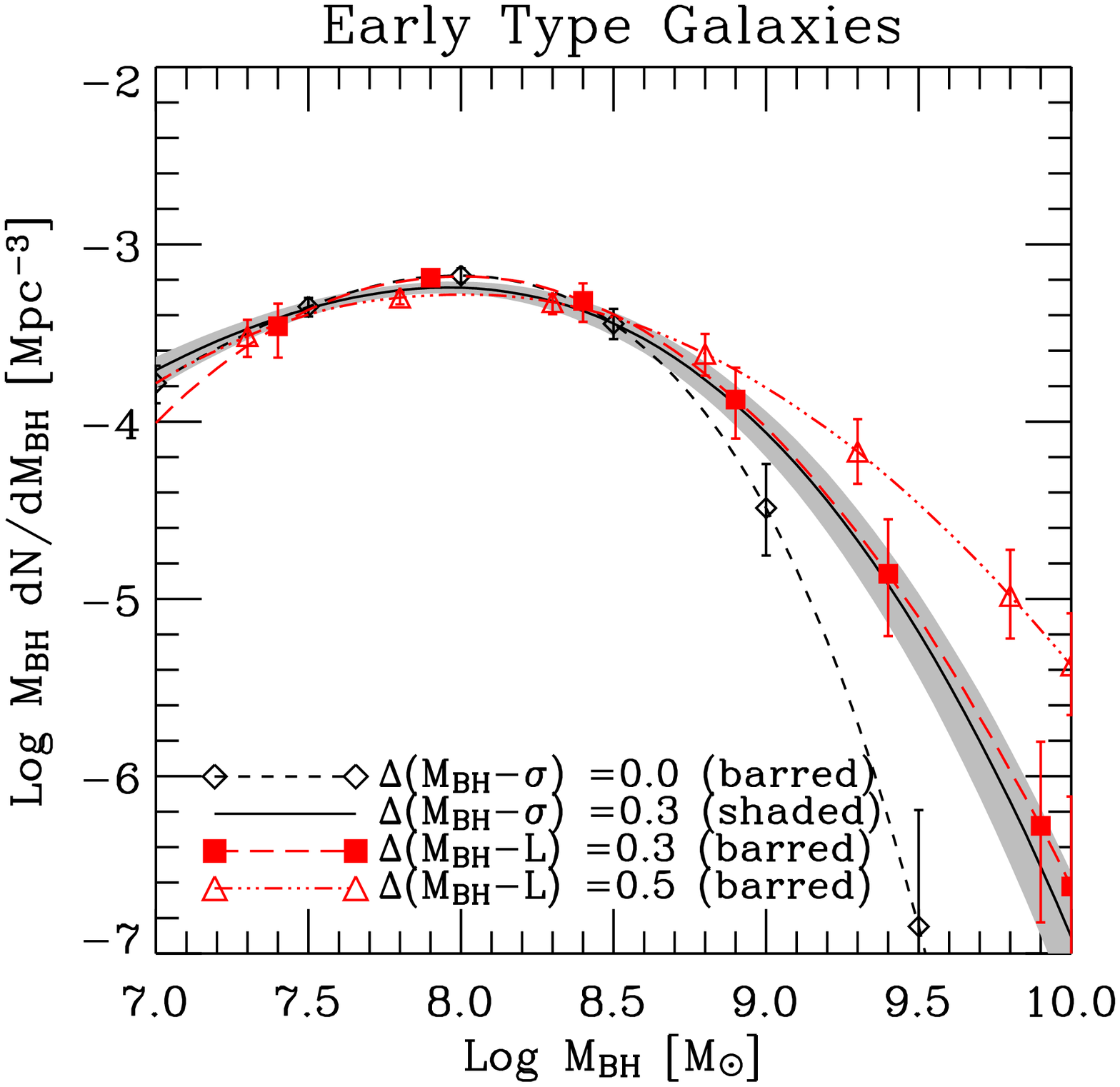}
\includegraphics[width=0.48\linewidth]{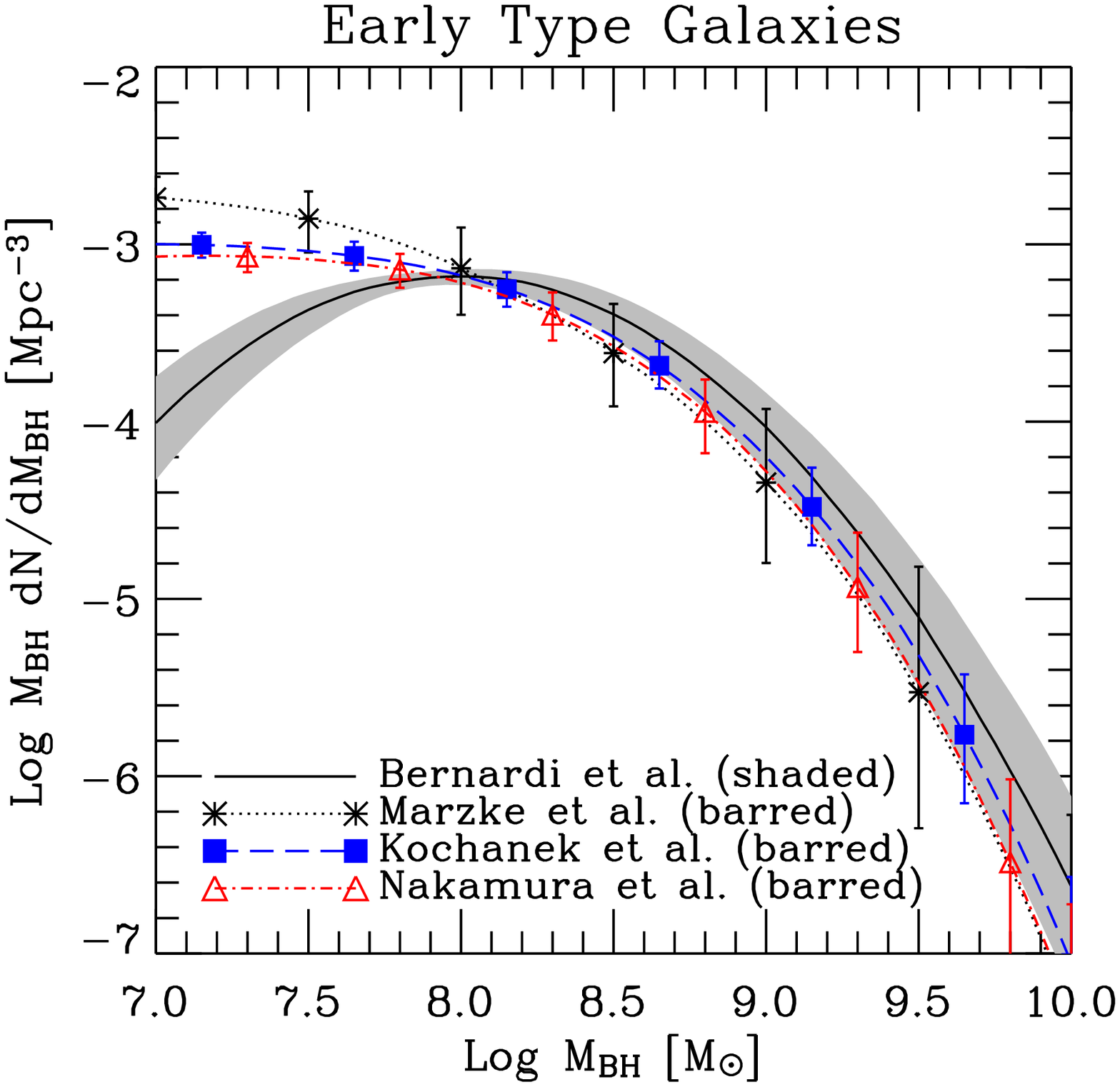}
\caption{\label{fig:bhearly}(a) Local BHMF for early type galaxies based on the
SDSS sample of \protect\cite{bernardisample}. The shaded area and error bars
("barred")
indicate the 16 and 84\%\ percentiles of 1000 Montecarlo realizations of the
local \BHMF\ and correspond to $\pm 1\sigma$ uncertainties (see text).  The
$\Delta$'s indicate the assumed intrinsic dispersions of the \MBHsig\ or
\MBHlum\ relations.  (b) Local BHMF for early type galaxies derived using the
luminosity functions from different surveys
(\protect\citealt{bernardilum,marzke,kochanek,nakamura}) and the \MBHlum\
relation with \sigBH=0.3.  The shaded area and errorbars have the same
meaning as in (a).}
\end{figure*}

We describe here the simple formalism which is commonly used to derive
the \BHMF\ from galaxy luminosity or velocity functions.

We define $\phi(x) dx$ as the number of galaxies per unit comoving volume with
observable $x$ (e.g.~luminosity $L$, or stellar velocity dispersion $\sigma$)
in the range $x, x+dx$.  The observable $y$ (e.g.~\MBH, the \BH\ mass) is
related to $x$ through the log-linear relation $\log y=a+b \log x$ and
$\Delta(y)$ is the intrinsic dispersion in $\log y$ at constant $\log x$.
Assuming a log-normal distribution then
\begin{equation}
P(\log y|\log x) = \frac{1}{\sqrt{2\pi}\Delta_y}\,
\exp\left[-\frac{1}{2} \left( \frac{\log y-a-b\log x}{\Delta(y)}\right)^2\right]
\end{equation}
where $P(\log y|\log x)d\log y$ is the probability that $y$ is in the
range $\log y,\log y+d\log y$ for a given $\log x$.
Thus the number of galaxies with $x,y$ in the ranges $x,x+dx$ and $y,y+dy$
is
\begin{equation}
\Phi(y,x)\,dx\,dy = \frac{P(\log y|\log x)}{y \ln 10} dy\, \phi(x)dx 
\end{equation}
$\Phi(y)dy$, the number of galaxies with $y$ in the range $y,y+dy$, is thus
the convolution of $\phi$ and $P$:
\begin{equation}
\label{eq:BHMF}
\Phi(y) = \int_0^{+\infty}
\frac{ P(\log y|\log x)}{y \ln 10}\,\phi(x)\,dx
\end{equation}
In the limit of zero-intrinsic dispersion:
\begin{equation}
\Phi(y) = \frac{x^{1-b}\phi(x) }{10^a\,b} 
\end{equation}
with $\log y=a+b \log x$.
After substituting $y$ with \MBH\ and, for instance, $x$ with $L$, the
spheroid luminosity, the total mass density in \BH's is simply
\begin{eqnarray}
\label{eq:rhoBH}
\lefteqn{\rhoBH} & ~~~~= & \int_0^{+\infty} M\,\Phi(M)\,dM = \nonumber\\
         &  & \int_0^{+\infty} dM \,
\int_0^{+\infty} dL \,
\frac{1}{\ln 10} P(\log M|\log L) \phi(L)
\end{eqnarray}
and in the zero intrinsic dispersion case:
\begin{equation}
(\rhoBH)_0 = 10^a\, \int_0^{+\infty} L^b \phi(L)\,dL
\end{equation}
where $M=\MBH$ for simplicity.

Inverting the order of integration in Eq.~\ref{eq:rhoBH}
and integrating on \MBH\ one
finally gets
\begin{equation}\label{eq:rhodisp}
{\rhoBH} = \mathrm{e}^{\frac{1}{2}[\sigMBH \ln 10]^2 } {\rhoBH}_0 
\end{equation}
A non-null \sigMBH\ increases \rhoBH\ by a factor
$\exp[\frac{1}{2}(\sigMBH \ln 10)^2 ]$ with respect to the zero dispersion
case, a fact already noted by \cite{yu02}.

\subsection{Consistency of the \MBHsig\ and \MBHlum\ Relations}
\label{sec:consistency}
Several authors (e.g.~\citealt{salucci, MS02, yu02, ferrarese, aller})
adopted the method just described
to determine the mass function of local \BH's. In the most recent works
the \MBHsig\ correlation has been preferred to \MBHlum\ on the ground that it
is tighter and, moreover, \cite{yu02} found a factor 2 discrepancy between the
values of \rhoBH\ determined by applying the \MBHsig\ and \MBHlum\ relations.
Thus, we first investigate this inconsistency of the \MBHsig\ and
\MBHlum\ relations by examining the \BHMF's derived by applying
the two relations to the velocity and luminosity functions obtained from the
same sample of galaxies.

We consider the SDSS sample of 9000 early type galaxies from \citealt{bernardisample} for which the luminosity and velocity functions were determined
independently (\citealt{bernardilum, sheth}). By `independently' we
mean that the velocity function was derived by directly measuring \sigstar\ of
all galaxies and not by applying the Faber-Jackson relation (hereafter FJ) to
the galaxy luminosity function (e.g. \citealt{gonzalez, aller, ferrarese}).
\cite{sheth} compared their velocity function
with that derived applying the FJ relation to the luminosity function and found
that it is fundamental to take into account the intrinsic dispersion of the
relation in order to obtain the correct velocity function.  It is consequently
expected
that the same might apply to the determination of the \BHMF.

To derive the \BHMF\ using the \MBHsig\ relation we apply Eq.~\ref{eq:BHMF}.
For the galaxy velocity function $\phi$
we use the parameterized form  for early
type galaxies by \cite{sheth}.
For the \MBHsig\ relation we use the coefficients obtained by fitting the
`Group 1' galaxies of \citealt{MH03} (i.e.~galaxies with
`secure' BH mass determinations):
\begin{equation}
\log\MBH = (8.30\pm0.07)+(4.11\pm0.33)(\log\sigstar-2.3)
\end{equation}
The slope is in agreement with that of \citealt{tremaine} ($4.02\pm0.32$)
but there is a larger normalization ($8.13\pm0.06$ i.e.~$\sim 0.2$ in \lMBH).
This choice of the coefficients for the \MBHsig\ relation
is made for consistency
since, in the following, we use the
coefficients of the \MBHlum\ relation determined for the same
'Group 1' galaxies by \cite{MH03}.
The local \BHMF's derived with the \MBHsig\ relation
assuming intrinsic dispersions
\sigMBH=0 and \sigMBH=0.30 
are shown in Fig.~\ref{fig:bhearly}a.
The shaded area and errorbars indicate the 16 and 84\%\ percentiles
of 1000 Montecarlo realizations of the local \BHMF. The realizations
were obtained by randomly
varying the input parameters assuming that they are normally distributed
with 1$\sigma$ uncertainties given by their measurement errors.
For the velocity function we have considered only the error on $\phi^\star$,
the number density at $\sigma^\star$, since the other errors are strongly
correlated among themselves (\citealt{sheth}).
The 16 and 84\%\ percentiles indicate the $\pm1\sigma$ uncertainties
on the logarithm of the local \BHMF, whose values from the Montecarlo
realizations are normally
distributed at a given \MBH.

To derive the \BHMF\ using the \MBHlum\ relation we again apply Eq.~\ref{eq:BHMF}.  The galaxy luminosity function
$\phi$ by \cite{bernardilum} is given as
a function of the total galaxy light. Since the
correlation of \MBH\ is with bulge light, we need to apply 
a correction $\Delta m$ in the case of S0 galaxies
to transform from total to bulge luminosity.
Following \cite{yu02}, the luminosity
function for the bulges of S0 galaxies is directly given by
\begin{equation}
\phi_b(m) = \frac{f_\mathrm{S0}}{f_\mathrm{E}+f_\mathrm{S0}}\phi(m-\Delta m)
\end{equation}
where $\phi$ is the luminosity function of early type galaxies,
$m$ is the absolute bulge magnitude, $\Delta m = m-m_\mathrm{total}$ and
$f_\mathrm{E}$, $f_\mathrm{S0}$ are the fractions of E and S0 galaxies with
respect to the total galaxy population.  The galaxy type fractions and $\Delta
m$ used in this paper are shown in Table \ref{tab:galtypes}.  The morphological
type fractions are from \citealt{fukugita98} (their Sbc fraction has been
evenly split between Sab and Scd). The $1\sigma$ uncertainties are a
conservative estimate we made after comparing various determinations of the
morphological type fractions available in the literature (see the discussion in
\citealt{fukugita98}).
The $\Delta m$ for the $B$ band are those estimated by \cite{aller} by
rebinning the \citealt{simien} data in the appropriate bins of galaxy types.
The $\Delta m$ for the $H$ band are based on data from \citealt{hunt04} and the
values for S0 galaxies were taken from the analysis by \cite{MH03}.
The $\Delta m$ values are little dependent on the
photometrical band at least within the considerable scatter, therefore in the
following analysis we will always use the $B$ band data 
regardless of the photometric band in which $\phi$ was measured.
\begin{table*}
 \centering
 \begin{minipage}{0.7\linewidth}
  \caption{Mass density of local \BH's for early and all galaxy types
  (Col.~3 and 4, respectively). Col.~1 indicates the galaxy luminosity
  or velocity function which was combined with the \MBH--host-galaxy-property
  in Col.~2 to determine the local \BHMF. All densities are computed with
  \sigBH=0.3. Uncertainties are the 16 and 84\%
  percentiles of the Montecarlo realizations described in the text. These
  correspond to $\pm 1\sigma$ uncertainties.}
  \label{tab:bhdens}
  \begin{tabular}{@{}lllllr@{}}
  \hline\hline
  Adopted Luminosity  & \MBH-- & \rhoBH\ (E+S0) & \rhoBH\ (All) \\
  or Velocity Function&  --host-gal.-prop.        & \multicolumn{2}{c}{[\rhoBHunits]} \\
  \hline
  Sheth et al. 2003    &\MBH-\sigstar &$3.4^{+0.6}_{-0.5}$&$5.0^{+1.7}_{-1.1}$\\
  Bernardi et al. 2003 &\MBH-\LbulK   &$3.8^{+1.2}_{-1.0}$&\dots              \\
  Nakamura et al. 2003 &\MBH-\LbulK   &$2.5^{+1.2}_{-0.8}$&$4.4^{+2.0}_{-1.3}$\\
  Kochanek et al. 2001 &\MBH-\LbulK   &$3.3^{+1.0}_{-0.7}$&$4.5^{+1.4}_{-1.1}$\\
  Marzke et al. 1994   &\MBH-\LbulK   &$3.4^{+2.7}_{-1.5}$&$4.5^{+3.1}_{-1.7}$\\
  \hline
  All L- or V- Func.     & \dots & \dots                  &$4.6^{+1.9}_{-1.4}$\\
  \hline
  \hline
 \end{tabular}
\end{minipage}
\end{table*}
Applying the above corrections to the early-type \rs\ band luminosity function
by \cite{bernardilum} and the color transformation
$\rs-K=2.8\pm0.2$		
we obtain the luminosity function of
the bulges of early types in the $K$ band. \cite{cole}
estimate $K=\zs-2.12$ which, combined with the average value for the early type
galaxies $\rs-\zs=0.68$ (\citealt{bernardilum}), provides the required color.
\cite{fukugita95} estimate $\rs-\zs=0.79:0.63:0.70:0.65:0.57$
(E:S0:Sab:Sbc:Scd) and this justifies our conservative 
choice of the scatter which also allows us to apply
the same color correction to all morphological types.
For the \MBHlum\ relation we use the coefficients obtained by fitting the
'Group 1' galaxies of \cite{MH03} in the $K$ band:
\begin{equation}
\log\MBH=(8.21\pm0.07)+(1.13\pm0.12)(\log\LKsph-10.9)
\end{equation}
where \LKsph\ is in units of \LKsun.
The local \BHMF\ derived with
the \MBHlum\ relation assuming intrinsic dispersions
\sigMBH=0.30 and \sigMBH=0.5
are plotted in \ref{fig:bhearly}a.
As before, the 16 and 84\%\ percentiles of 1000 Montecarlo realization of the
local \BHMF\ indicate $\pm 1\sigma$ uncertainties (shaded area and errorbars).

As clear from Fig.~\ref{fig:bhearly}a the effect of including the intrinsic
dispersion in the \MBHsig\ and \MBHlum\ correlations is that of softening the
high mass decrease of the \BHMF, thus increasing the total density. But the
most important result is that in order to provide the same \BHMF, the relations
\MBHsig\ and \MBHlum\ must have the same intrinsic dispersion to within 0.1 in
log.  \cite{yu02} found that $\rhoBH[\MBHlum] \sim 2\rhoBH[\MBHsig]$ by
using \sigMBH=0 for  \MBHsig\ and \sigMBH=0.5 for \MBHlum.  This discrepancy 
can be
entirely ascribed to the effect quantified by Eq.~\ref{eq:rhodisp}. Indeed, the
densities \rhoBH\ of the \BHMF's plotted in Fig.~\ref{fig:bhearly}a are
$\rhoBH=2.7^{+0.5}_{-0.4}\rhoBHunits$ [\MBHsig\ with \sigMBH=0, in agreement
with \cite{yu02}] and $\rhoBH=5.5_{-1.3}^{+2.0}\rhoBHunits$ [\MBHlum\ with
\sigMBH=0.5].  Conversely, with \sigMBH=0.30, one obtains
$3.4^{+0.6}_{-0.5}\rhoBHunits$ (\MBHsig) and $3.8^{+1.2}_{-1.0}\rhoBHunits$
(\MBHlum) and these two values are in excellent agreement.
The densities in massive \BH's were evaluated in the $\log(\MBH/\Msun)=6-10$
range and the same range will be considered throughout the
rest of the paper.

One advantage of using the \MBHsig\ relation is that the uncertainties on the
derived \BHMF\ are smaller than in the case of \MBHlum. This is because
with the \MBHsig\ relation one does not have to apply any correction for
the bulge fraction.  On the other hand, measuring stellar velocity dispersions
is much more difficult
than measuring galaxy luminosities; thus it is clear that the
two relations \MBHsig\ and \MBHlum\ should complement each other.

In summary, the use of the same intrinsic dispersion for the \MBHsig\ and
\MBHlum\ relations provides perfectly consistent \BHMF's with the same mass
densities \rhoBH.  This is a confirmation of the results by \cite{MH03}
who showed that, when considering only secure BH measurements, the
\MBHsig\, \MBHlum\ and \MBH-\Msph\ relations have similar intrinsic dispersions
($\sim 0.3$ in log).  When not taking into account the intrinsic dispersion of
the \MBHsig\ relation, the local \BHMF\ is systematically underestimated at the
high mass end, where the disagreement with the \BHMF\ of AGN
relics has been claimed.
\begin{figure*}
\centering
\includegraphics[width=0.48\linewidth]{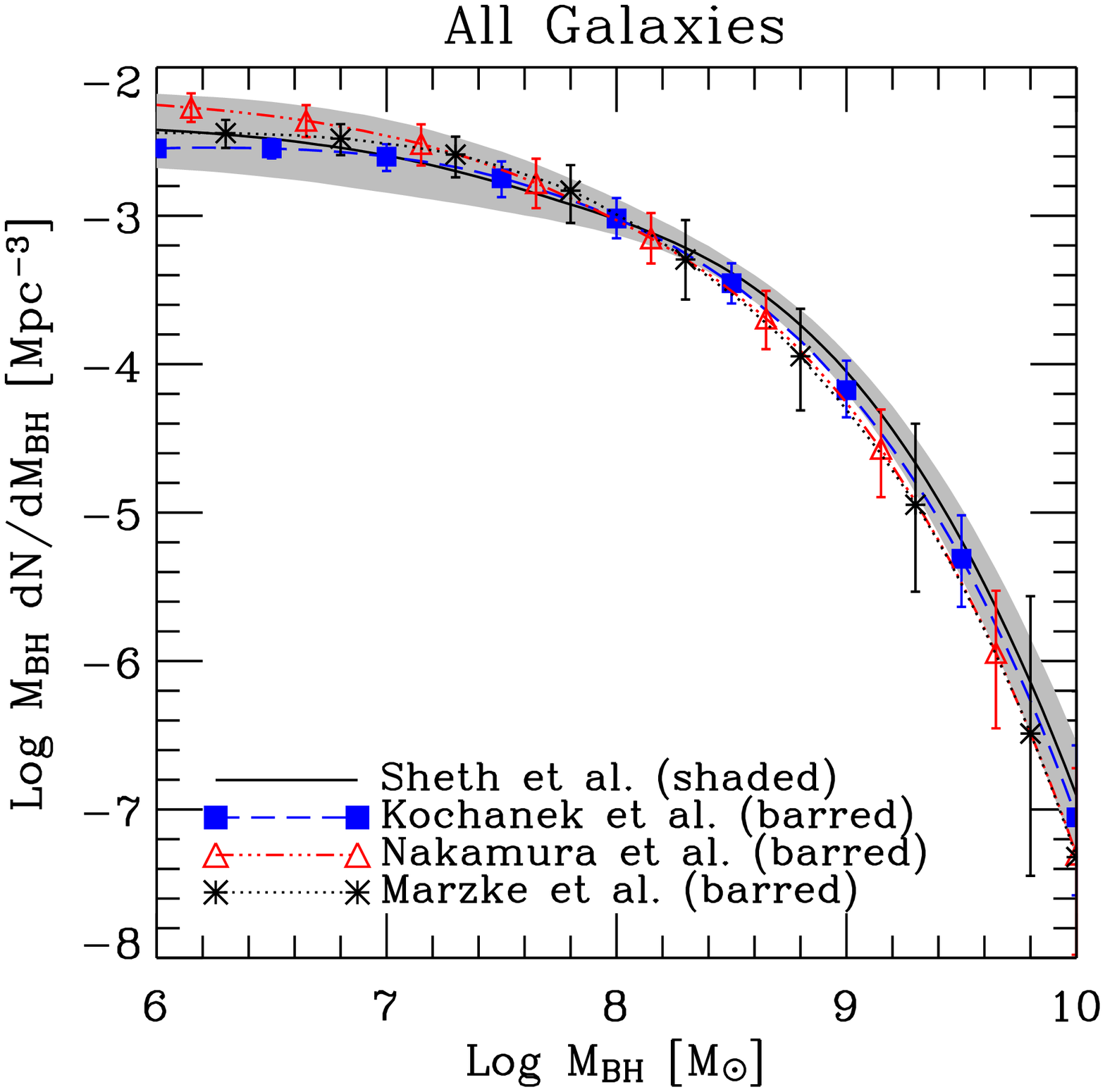}
\includegraphics[width=0.48\linewidth]{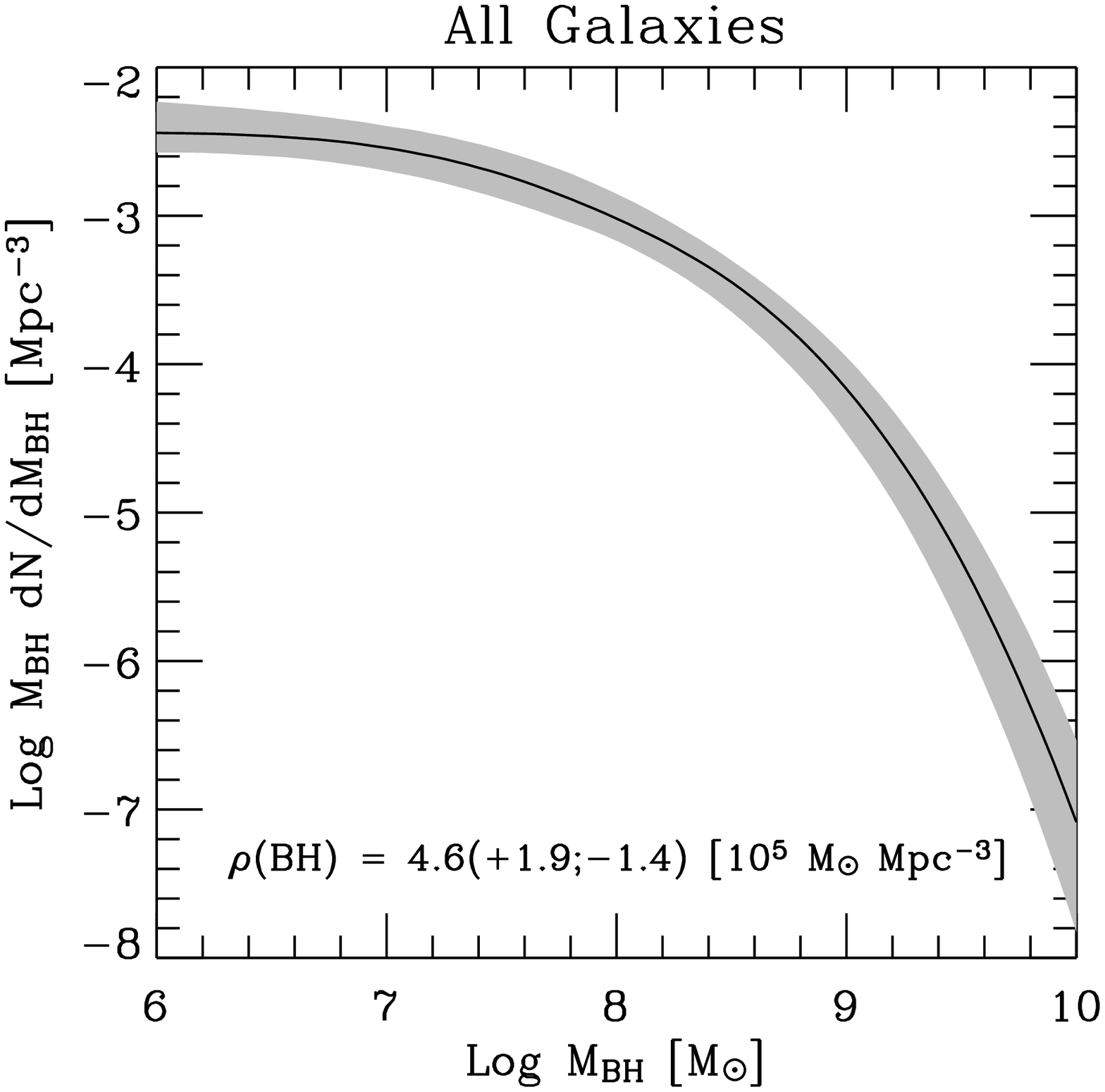}
\caption{\label{fig:bhall} (a) Local \BHMF's for all galaxy types derived using
different galaxy surveys and shown with the same notation as in
Fig.~\protect\ref{fig:bhearly}a.  (b) Best estimate of the local \BHMF\
obtained by combining all the Montecarlo realizations of the \BHMF\ which were
used for panel (a). The solid line represents the 50\%\ percentile and
the shaded area is delimited by the 16 and 84\%\ percentile levels.}
\end{figure*}

\subsection{The Black Hole Mass Function for Early Type Galaxies}

Having established that both \MBHsig\ and \MBHlum\ relations provide consistent
\BHMF's we can now evaluate the effects of using luminosity functions from
different galaxy surveys and photometric bands in determining the \BHMF\ in
early type galaxies.

In Fig.~\ref{fig:bhearly}b we compare the local \BHMF's for early type
galaxies obtained from different galaxy luminosity functions, in different
photometric bands. We use the luminosity functions by
\cite{bernardilum}, \cite{marzke}, \cite{kochanek}, and \cite{nakamura}
with details of the derivation specified in the following.
\begin{itemize}
\item \citealt{bernardilum}: this is the same \BHMF\ plotted in panel a
derived using the \MBHlum\ relation.
\item \citealt{marzke}: we use the luminosity functions per morphological
type from the CfA survey. The luminosities are in Zwicky
magnitudes, $M_Z$, and we apply the color transformations directly measured by
\cite{kochanek} computing $M_Z-K$ for all the objects used for the
luminosity functions ($K$ is obtained from the 2MASS catalogue):
$M_Z-K=(4.1\pm0.65; 3.95\pm0.65; 3.79\pm0.56; 3.34\pm0.64)$ for
(E;S0;Sa-Sb;Sc-Sd).  The $K$ magnitudes used are isophotal magnitudes $K_{20}$
and the correction to total magnitudes is $K_{tot}=K_{20}-(0.2\pm0.04)$
(\citealt{kochanek}).  Then we use the bulge-total correction and the
\MBH-\LKsph\ relation from Tab.~\ref{tab:galtypes}.
\item \citealt{kochanek}: we use the luminosity function in the K band
and we apply the correction from $K_{20}$ to $K_{tot}$ described
previously, and we use the morphological fractions and Bulge-Total corrections
from Tab.~\ref{tab:galtypes}.
\item \citealt{nakamura}: we use the luminosity function in the SDSS \rs\ band
and we apply the color correction $K =
\rs-(2.8\pm0.2)$ described previously and the Bulge-Total corrections of
Tab.~\ref{tab:galtypes}.
\end{itemize}
As in the previous section,  uncertainties are estimated with 1000 Montecarlo
realizations of the local \BHMF\ where, in the case of the galaxy luminosity
functions, we use only errors on $\Phi^\star$, the galaxy number density.  All
the local \BHMF's for early type galaxies are in remarkable agreement within
the uncertainties.
The discrepancy at the low mass end ($\MBH<\ten{8}\Msun$) between the \BHMF\
derived with the \cite{bernardilum} luminosity functions and the others is not
significant. It occurs in a region where the luminosity functions of early
type galaxies are extrapolated and is due to the different functional forms
adopted to fit the data (gaussian for \cite{bernardilum}, Schechter functions
for the others).

\subsection{The Black Hole Mass Function for All Galaxy Types}
\label{sec:localbhmf}

Here we estimate the local \BHMF\ by considering also late morphological
types. We use all the luminosity and velocity functions described in the previous section (with
the exception of the one by \cite{bernardilum} which is only for early type
galaxies). For the velocity function of late type galaxies we take
the estimate made by \cite{sheth} and shown in their Fig.~6.
When necessary, we apply the bulge-to-total correction with the numbers
in Tab.~\ref{tab:galtypes} and we then use either the \MBHlum\ or the \MBHsig\
relation as described in Sec.~\ref{sec:consistency}.

Fig.~\ref{fig:bhall}a shows the local \BHMF\ for all galaxies while
Tab.~\ref{tab:bhdens} reports the estimated \BH\ mass densities \rhoBH\ both for
early type and all galaxies. These densities are computed for $h=0.7$.  All the
\BHMF's and \rhoBH's are in agreement within the errors.  Finally, in
Fig.~\ref{fig:bhall}b we present our best estimate of the local \BHMF\ obtained
by merging all the random realizations of the \BHMF's shown in
Fig.~\ref{fig:bhall}a and considering the 16, 50 and 84\% percentile levels.
Our best estimate of the local density in massive BH's is consequently
$\rhoBH=4.6^{+1.9}_{-1.4}\rhoBHunits$. Roughly 70\%\ of the total BH density
resides in early type galaxies.  Our estimate of the density in local \BH's is
in agreement with \cite{merritt01} and with \cite{ferrarese} though, in the
latter case, the shape of our \BHMF\ is very different at the high mass end.
Our estimate is a factor $\simeq 1.8$ larger than those by \cite{yu02} and
\cite{aller} and the reasons for this discrepancy are outlined in
Sec.~\ref{sec:localrelics}.

\section{The Mass Function of AGN relics}
\label{sec:rbhmf}

Once the mass function of local \BH's has been determined, the subsequent step
is the estimate of the \BHMF\ of AGN relics, i.e.~\BH's in galactic nuclei
which were grown exclusively during active phases from small (1-\ten{3}\Msun)
seeds. We will first describe the continuity equation which will be used to
relate the relic \BHMF, $N(M,t)$, to the AGN luminosity function $\phi(L,z)$,
under the assumption that AGN are powered by mass accretion onto massive \BH's.
The physical quantity directly related to the mass accretion onto the BH is the
total {\it intrinsic} AGN luminosity $L$.
Since AGN luminosity functions are determined in
limited energy bands, we will provide suitable bolometric corrections. We will
then briefly describe the adopted AGN luminosity functions obtained with our
bolometric corrections.  Finally we will present our estimates of the relic
BHMF's derived from different AGN luminosity functions.

\subsection{The Continuity Equation}
\label{sec:conteq}

The Black Hole Mass Function \BHMF\ of AGN relics, $N(M,t)$, can be estimated
from the continuity equation (\citealt{cavaliere, small})
with which, under simple assumptions, it is possible to relate $N(M,t)$ to the
AGN luminosity function.  If $\NBH \mathrm{d}M$ is the comoving number density
of \BH's with mass in the range $M$ and $M+\mathrm{d}M$ at cosmic time $t$, the
continuity equation can be written as:
\begin{equation}\label{eq:continuity}
\frac{\partial\NBH}{\partial t} +
\frac{\partial}{\partial M} \left[\NBH \langle\MdotBH\rangle \right] = 0
\end{equation}
where $\langle\MdotBH\rangle$ is the "average" accretion rate on the BH of mass $M$.
We adopt the working assumption that AGN's are powered by accretion onto \BH's,
and that the \BH\ growth takes place during phases in which the AGN is
accreting at a fraction $\lambda$ of the Eddington limit ($L=\lambda\LEDD$)
converting mass into energy with an efficiency $\eff$.  We can thus
simply relate the AGN Luminosity Function \NAGN\ [$\NAGN \mathrm{dlog} L$ is
the comoving number density of AGNs in the range $\logL, \logL
+\mathrm{d}\logL$ at cosmic time $t$] to the \BH\ mass function:
\begin{equation}\label{eq:phiagn}
\phi(L,t) \dlogL = \delta(M,t) N(M,t) \dM
\end{equation}
where $\delta(M,t)$ is the fraction of \BH's with mass $M$ which are active at
time $t$, i.e.~the \BH\ duty cycle.
If a \BH\ is accreting at a fraction $\lambda$ of the Eddington
rate, its emitted luminosity is
\begin{equation}\label{eq:MbhL}
L=\lambda \frac{Mc^2}{t_E} = \eff \Mdot_\mathrm{acc} c^2
\end{equation}
where $t_E$ is the Eddington time, $\eff$ is the accretion efficiency
and $\Mdot_\mathrm{acc}$ is the matter falling onto the black hole. The growth
rate of the \BH, \Mdot, is thus given by $\Mdot =(1-\eff)
\Mdot_\mathrm{acc}$, since a fraction $\eff$ of the accreted mass is
converted into energy and thus escapes the BH.  Since $\langle\MdotBH\rangle =
\delta(M,t) \MdotBH$, combining Eqs. \ref{eq:phiagn} and \ref{eq:MbhL} we can
write:
\begin{equation}\label{eq:mdot}
\NBH \langle\MdotBH\rangle = \frac{(1-\eff)}{\eff c^2 \ln 10} 
\left[ \phi(L,t) \right]_{L=\lambda \frac{Mc^2}{t_E}} \frac{dL}{dM}
\end{equation}
which can be placed in Eq.~\ref{eq:continuity} where the only unknown function
is $N(M,t)$.  If $\eff$ and $\lambda$ are constant we can then write
\begin{equation}\label{eq:NBH}
\frac{\partial\NBH}{\partial t} = -\frac{(1-\eff)\lambda^2c^2}{\eff t_E^2 \ln 10}
\left[ \frac{\partial\phi(L,t)}{\partial L} \right]_{L=\lambda\frac{Mc^2}{t_E}}
\end{equation}
which can be easily integrated given the AGN luminosity function and the
initial conditions.

For initial condition, we assume that, at the starting redshift \zo,
$\delta[M,t(\zo)]=1$. This can be interpreted either by saying that at \zo\
all Black Holes are active or that we are following the evolution only of those
\BH's which were active at \zo.  Thus
\begin{equation}
M N(M,\tzo) =
\left[\phi(L,\tzo)\right]_{L=\lambda\frac{Mc^2}{t_E}}
\end{equation}
We will show that the final results are little sensitive to the choice of the
initial conditions, provided that $\zo > 3$, since most of the \BH\ growth
takes place at lower redshifts.

Eq.~\ref{eq:NBH}, which represents the continuity equation in the case of
constant $\eff$ and $\lambda$, can be trivially integrated on $M$ to
derive the relation used by various authors
(\citealt{padovani90,chokshi,yu02,ferrarese}):
\begin{equation}\label{eq:chokshi}
\rhoBH = \frac{1-\eff}{\eff c^2} \, U_{T}
\end{equation}
note the factor $(1-\eff)$ which is needed to account for the part 
of the accreting matter which is
radiated away during the accretion process.  $U_{T}$ is the total {\it comoving}
energy density from AGN's (not to be confused with the total {\it observed}
energy density) and is given by 
\begin{equation}\label{eq:Ecomov}
U_T = \int_0^{{\zo}} dz \frac{dt}{dz} \int_{L_1}^{L_2} L \phi(L,z)\,d\log L
\end{equation}
$\phi(L,z)\,d\log L$ is used if the AGN luminosity function is defined per
logarithmic luminosity bin, otherwise it should be $\phi(L,z)\,dL$.

The right-hand element of the continuity equation, which contains the source
function, is null, meaning that we neglect any process which, at time $t$, might
`create' or `destroy' a \BH\ with mass $M$. Indeed, in the merging process of
two \BH's, $M_1+M_2\rightarrow M_{12}$, \BH's with $M_1$ and $M_2$ are
destroyed while a BH with $M_{12}$ is created.  We decided to neglect merging
of \BH's because, at present, the merging rate, $\gamma(M_1,M_2,z)$, is very
uncertain and dependent on the assumptions of the model with which it is
computed.  Moreover, the aim of this paper is to assess if local \BH's are
indeed relics of AGN activity and by neglecting merging we can assess the
importance of mass accretion during AGN phases.  \cite{yu02} have used an
integral version of the continuity equation in order to take into account
merging but their approach does not allow a direct comparison of the local and
relic \BHMF's. Furthermore, the discrepancy they find between local \BH's and
AGN relics becomes worse if merging is important. \cite{granato04} in their
physical model of coevolution of \BH\ and host galaxy find that the growth of
the \BH\ is due to mass accretion. Similarly, \cite{haiman03} find that merging
of BH's is important only at high redshifts while, at lower $z$, BH growth is
dominated by mass accretion.  These reasons support our choice of neglecting
the merging process.  Indeed merging of \BH's might well be very important in
shaping the \BHMF\ at high redshifts ($z>3-5$) but, as we will see, the relic
\BHMF\ at $z=0$ is very little dependent on its shape and normalization at
$z\sim 3$. Moreover, if the relic \BHMF\ at $z=0$ is changed significantly by
merging then its remarkable agreement with the local \BHMF\ in the merging-free case would be a mysterious coincidence (see
Sec.~\ref{sec:localrelics}).

\subsection{The Bolometric Corrections}
\begin{figure*}
\centering
\includegraphics[width=0.48\linewidth]{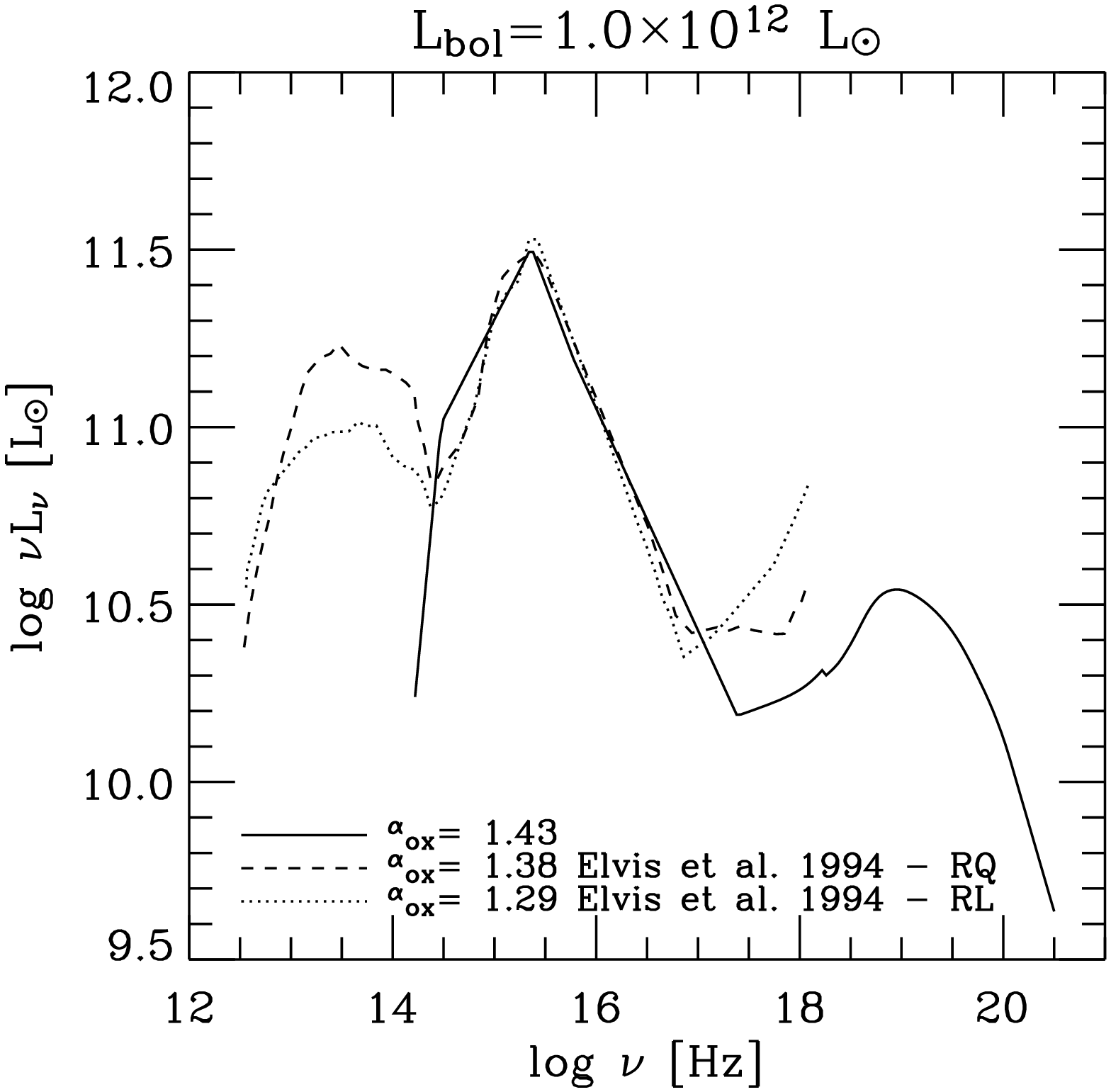}
\includegraphics[width=0.48\linewidth]{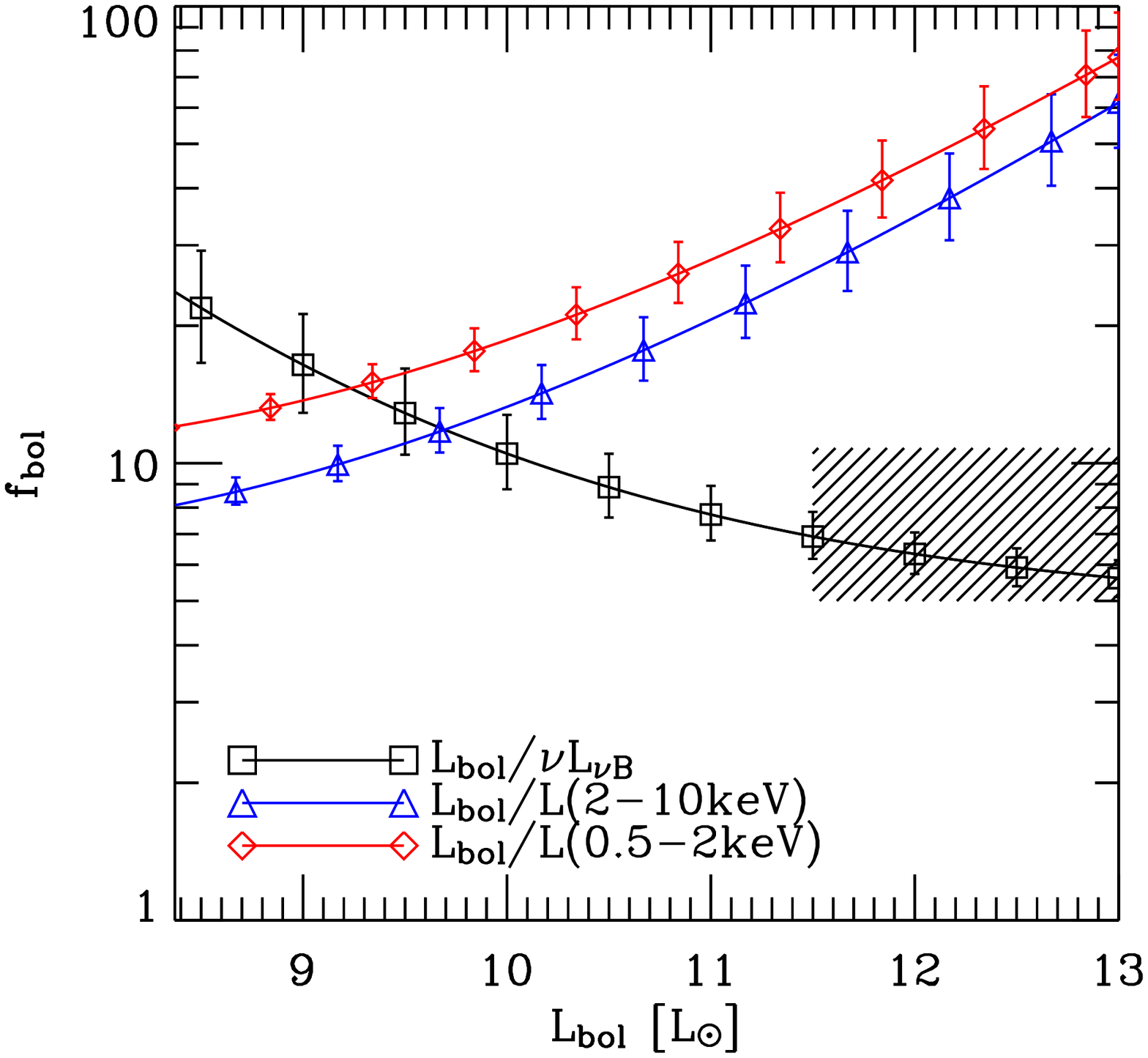}
\caption{\label{fig:specbolcorr}
(a) Template spectrum for a $L=\ten{12}\Lsun$ AGN obtained as described in the
text (solid line).
The dashed and dotted lines represent the radio quiet and radio loud
templates by Elvis et al. 1994.  (b) Bolometric corrections derived from the
spectral templates built as described in the text.  The error bars represent
the 16 and 84\% percentiles of the 1000 Montecarlo realizations described in
the text.  The hatched area represents the B band bolometric correction (with
$\pm 1 \sigma$ scatter) by Elvis et al. 1994.}
\end{figure*}

The \BH\ mass accretion rate is given by the AGN luminosity $L$ which, in turn,
can be estimated from the luminosity in a given band $b$, $L_b$, by applying a
suitable bolometric correction $f_{bol,b} = L/L_b$.  At the beginning of
Sec.~\ref{sec:rbhmf} we have referred to $L$ as the total {\it intrinsic} AGN
luminosity, and before presenting the adopted bolometric corrections it is
necessary to define the meaning of {\it intrinsic}.

The integral of the observed spectral energy distribution (SED) of an AGN
provides the total {\it observed} luminosity, $L_\mathrm{obs}$, thus 
the quasar bolometric corrections by \cite{elvis} provide
$L_\mathrm{obs}$ because they are based on the average {\it observed} SED.
However, $L_\mathrm{obs}$ does not give an accurate estimate of the BH mass
accretion rate because it often includes reprocessed radiation
(i.e.~radiation absorbed along other lines of sight and re-emitted
isotropically).  The accretion
rate is better related to the total luminosity {\it directly} produced by the
accretion process, which we call the total {\it intrinsic} luminosity $L$.  In
AGN's, $L$ is given by sum of the Optical-UV and X-ray luminosities radiated by
the accretion disk and hot corona, respectively. Conversely, it is well known
that the IR radiation is reprocessed from the UV (e.g.~\citealt{antonucci93}).
Thus, in order to estimate $L$ one has to remove the IR bump from the observed
SED's of unobscured AGN's.
In radio loud AGN's there might be an important contribution from
synchrotron radiation to the IR. However this will not affect our average
bolometric corrections since the number of radio loud AGN's is small, $\sim
10\%$ of the whole AGN population.

In previous works, several authors have used the bolometric corrections by
\cite{elvis} but, as explained above, these provide the {\it observed} and not
the {\it intrinsic} AGN luminosity.  Removing the IR bump from the 
\citealt{elvis} spectral templates, the 1\MIC-X-ray range includes
$\simeq 2/3$ of the total energy; thus one obtains, for instance, $L/\nu_B
L_{\nu B}=7.9\pm 2.9$ instead of the commonly used $L/\nu_B L_{\nu B}=11.8\pm
4.3$.  However, the \citealt{elvis} quasar
templates, obtained from an X-ray selected sample of quasars, are X-ray
``bright'' (see the discussion in \citealt{erz02}) and thus underestimate the
bolometric corrections in the X-ray energy bands. Moreover the dependence on
luminosity of the bolometric corrections cannot be obtained.  For these reasons
we will derive new estimates of the
bolometric corrections as described in the following.

First, we construct a template spectrum. In the optical-UV region our
template consists of a broken power-law with $\alpha=-0.44$ ($L_\nu\propto
\nu^\alpha$) in the range $1\MIC<\lambda<1300$\AA\ and $\alpha = -1.76$ in the range 1200-500\AA\ (e.g.~\citealt{telfer,vandenberk}). At
$\lambda > 1\MIC$ our big blue bump is truncated assuming $\alpha=2$ as in the
Rayleigh-Jeans tail of a blackbody.  The X-ray spectrum beyond 1 keV consists
of a single power law plus a reflection component. The powerlaw has the typical
photon
index $\Gamma = 1.9$ (e.g.~\citealt{george98,perola02}) and an exponential cutoff at
$E_c=500\KEV$.  Following \cite{ueda},
the reflection component is generated with the {\it pexrav}  
model (\citealt{pexrav}) in the XSPEC package with a 
reflection solid angle of $2\pi$,
inclination angle of $\cos i=0.5$ and solar abundances.
We then rescale the X-ray spectrum to a given $\alpha_{ox}$
(\citealt{zamorani81}) which is defined as
\begin{equation}
\alpha_{ox} = -\frac{\log[L_\nu(2500\mathrm{\AA})/L_\nu(2\KEV)]}
{\log[\nu(2500\mathrm{\AA})/\nu(2\KEV)]}
\end{equation}
and we finally connect with a simple powerlaw the point at 500\AA\ with that at
1\KEV. To account for the dependence of $\alpha_{ox}$ on luminosity, we use the
relation by \cite{vignali}:
\begin{equation}
\alpha_{ox}=-0.11*\log L_\nu(2500\mathrm{\AA})+1.85
\end{equation}
Finally, we assume that the template spectra,
hence the bolometric corrections, are independent of redshift. 

The template spectrum for a $L=\ten{12}\Lsun$ AGN is plotted in
Fig.~\ref{fig:specbolcorr}a and is compared with the radio quiet and radio loud
quasar median spectra by \cite{elvis}.  Our template is in very good
agreement in the optical-UV part but is obviously missing the IR bump, which
is due to reprocessed UV radiation and thus not considered here. Our
template is fainter in X-rays but this is expected since, as already discussed,
the quasar sample by \cite{elvis} is X-ray bright.

\begin{figure*}
\centering
\includegraphics[width=0.48\linewidth]{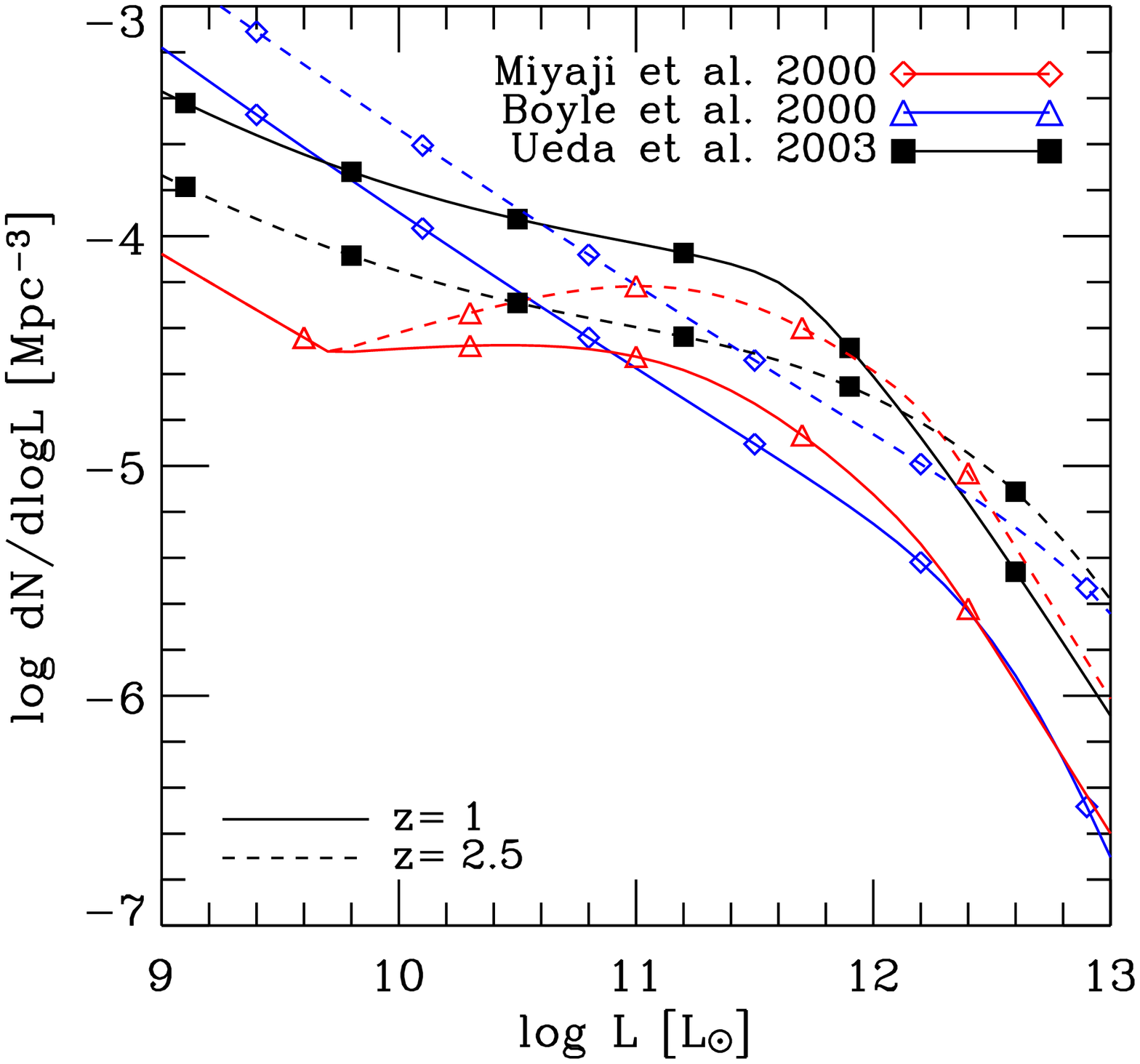}
\includegraphics[width=0.48\linewidth]{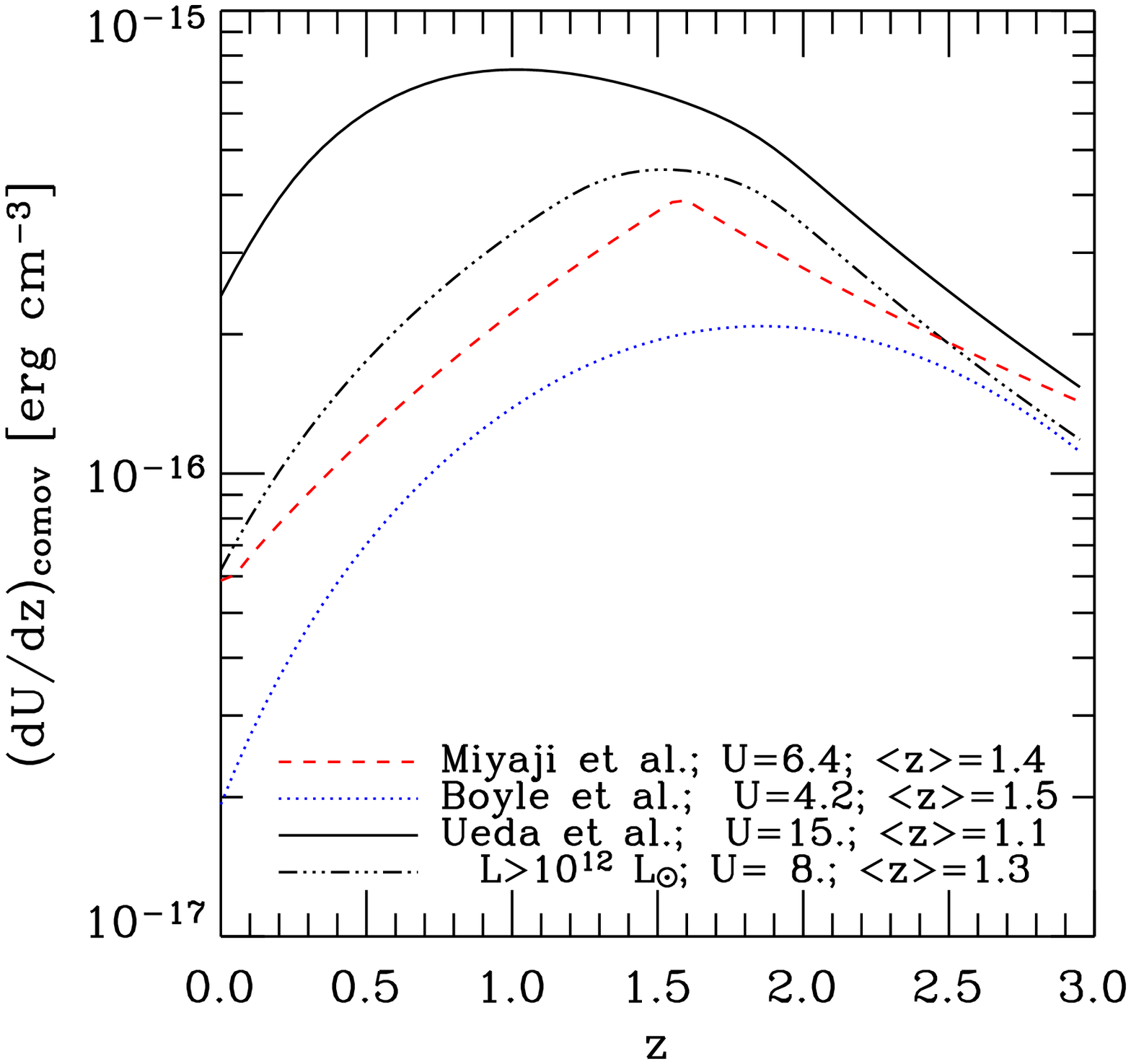}
\caption{\label{fig:agnlf}(a) AGN LF's at selected redshifts obtained with the
bolometric corrections described in the text. Different symbols
indicate different AGN luminosity functions, while different
line styles distinguish between 
redshifts. (b) Differential comoving energy
densities of AGN's from different surveys obtained with the bolometric
corrections described in the text. In the case of the \protect\cite{ueda}
luminosity function we also plot the comoving energy density for 
high luminosity objects ($L>10^{12}\Lsun$). 'U', the total comoving energy density in the redshift range $z=0-3$, is in units
of $10^{-16}\ERG\CM\3$ and '$<z>$' is the average redshift
computed following Eq.~\protect\ref{eq:zav}.}
\end{figure*}

In Fig.~\ref{fig:specbolcorr}b we plot the bolometric corrections
$L/\nu_BL_{\nu B}$, $L/L(0.5-2\KEV)$ and $L/L(2-10\KEV)$ derived
from the above templates.
The errorbars lines represent the 16 and 84\%\ percentiles from 1000
Montecarlo realization of the spectral templates where we have assumed
the following $1\sigma$ uncertainties for the input parameters:
$\pm0.1$ for the spectral slopes ($\alpha$; \citealt{telfer, vandenberk})
in the 1\MIC-500\AA\ range
and $\pm 0.05$ for the $\alpha_{ox}$, constant at all luminosities. The latter
is a conservative assumption (see, e.g., \citealt{yuan}) but is made to account
for possible, but unaccounted for, systematic errors.
As in the case of the local \BHMF, the 16 and 84\%\ percentiles
correspond to $\pm 1\sigma$ uncertainties on the logarithm of the
bolometric corrections.
The 50\%\ percentiles can be fit 
with a 3rd degree polynomial to obtain the following convenient relations:
\begin{eqnarray}
\lefteqn{\log(L/L(2\!-\!\!10\KEV))=1.54+0.24{\cal L}+0.012{\cal L}^2-0.0015{\cal L}^3} \nonumber\\
\lefteqn{\log(L/L(0.5\!-\!\!2\KEV))=1.65+0.22{\cal L}+0.012{\cal L}^2-0.0015{\cal L}^3} \nonumber\\
\lefteqn{\log(L/\nu_BL_{\nu B})=0.80-0.067{\cal L}+0.017{\cal L}^2-0.0023{\cal L}^3} 
\end{eqnarray}
where ${\cal L} = (\logL-12)$ and $L$ is the bolometric luminosity in
units of \Lsun. Hence, the bolometric corrections have log-normal distributions
with average values given by the above equations and $\pm1\sigma$ scatters (at
fixed $L$) which can be derived from Fig.~\ref{fig:specbolcorr}b.
Scatters are
given by $\sim 0.05$ for the B band and $\sim 0.1$ for the X-rays, taken
independent of $L$ for simplicity.  It is worth noting that the B band
bolometric correction is in agreement with that by \cite{elvis} whose average value and scatter are shown by the hatched
area in Fig.~\ref{fig:specbolcorr}b. The scatter
of the \cite{elvis} bolometric correction is the standard deviation of their 
quasar sample while ours are
uncertainties on the average values.
The average $\alpha_{ox}$ in the $\log(L/\Lsun) = 11.5-12.5$ range is 1.43, the
same value estimated by \cite{erz02} after correcting for the biases due to
X-ray or optical selection of the parent quasar samples.
\begin{figure*}
\centering
\includegraphics[width=0.48\linewidth]{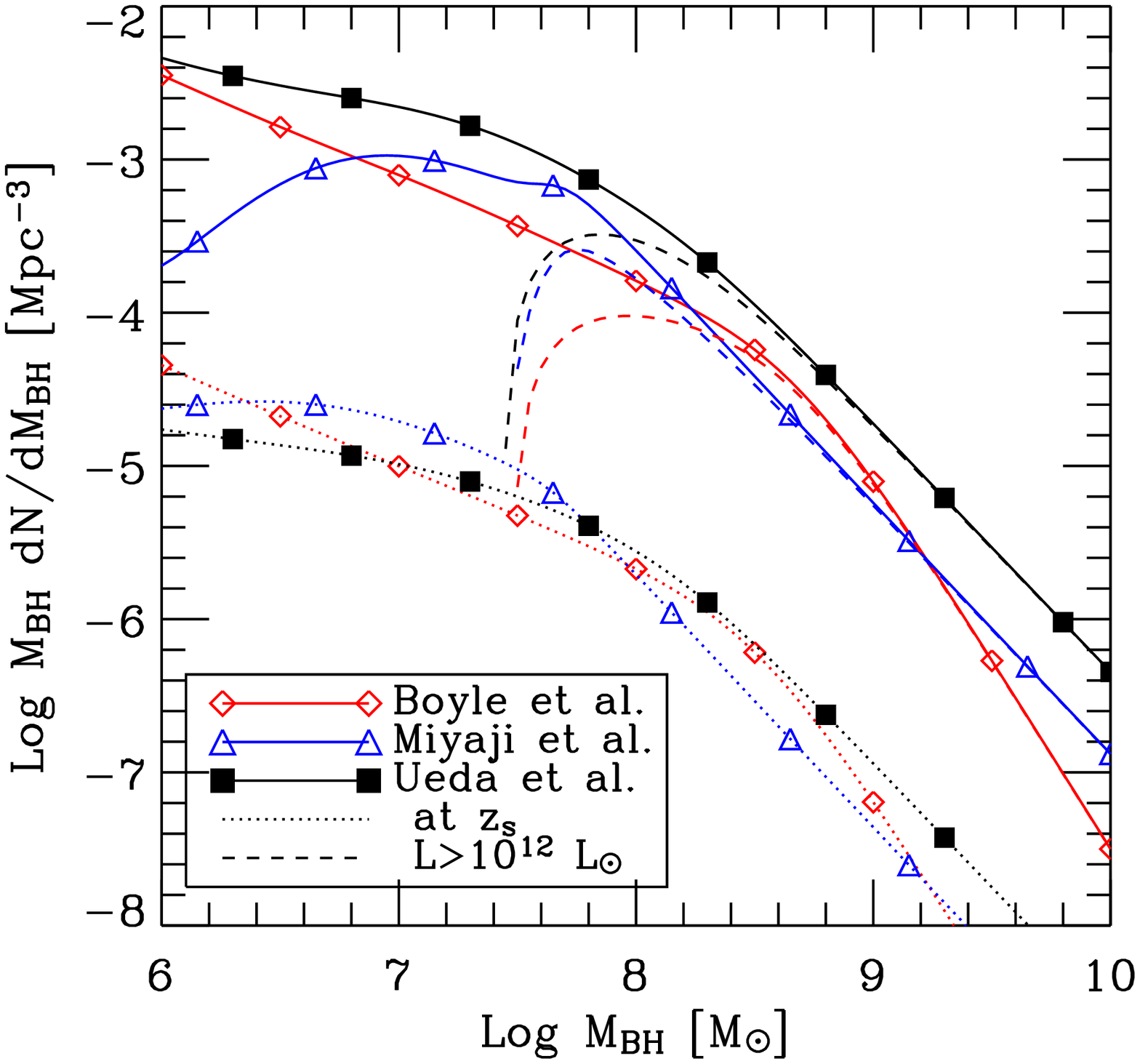}
\includegraphics[width=0.48\linewidth]{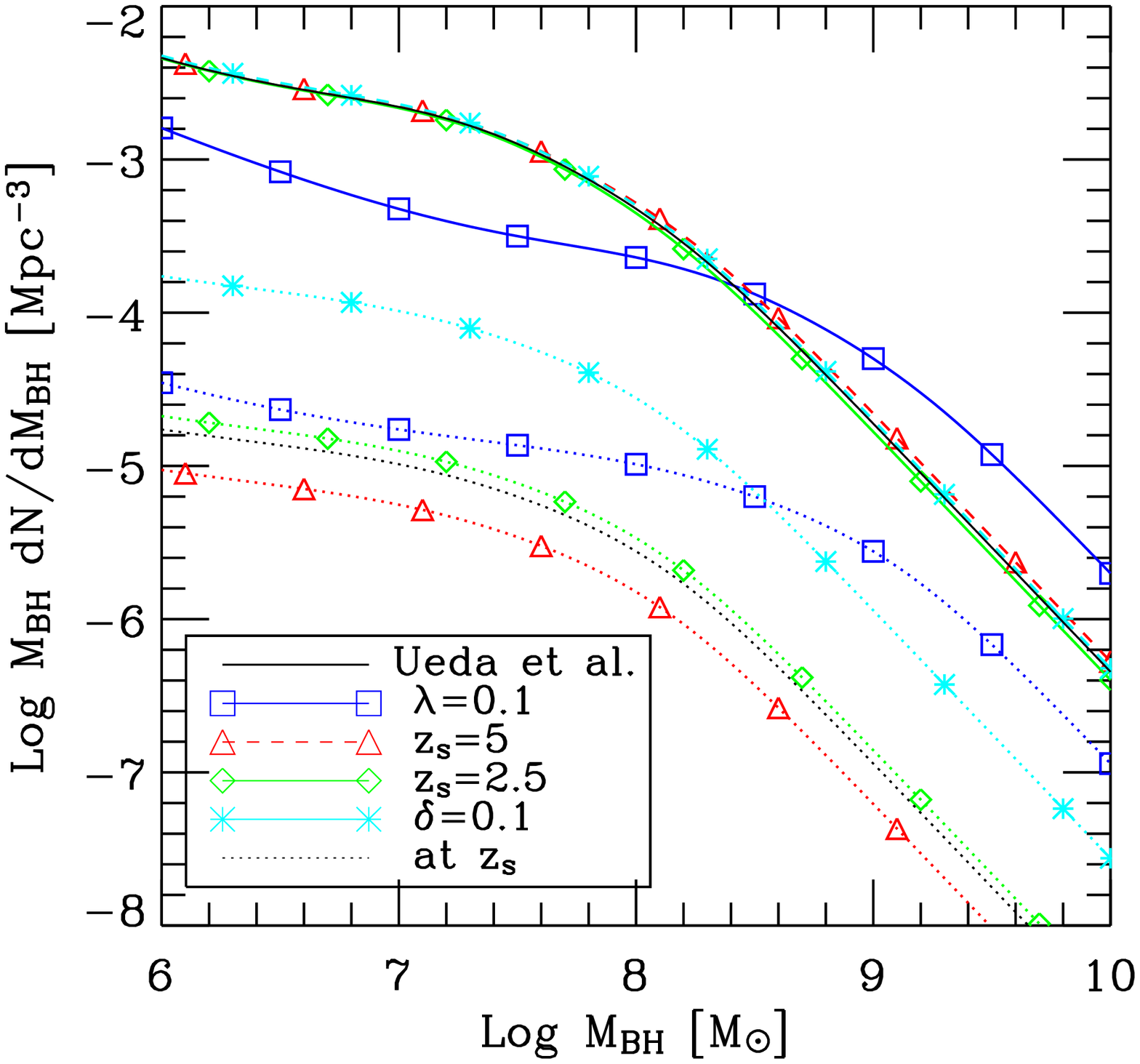}
\caption{\label{fig:relics1}
(a) Relic \BHMF's at $z=0$ (solid lines) calculated assuming $\eff=0.1$ and
$\lambda=1$ compared with the relic \BHMF's at \zo=3 (dotted lines). Different
symbols indicate the different AGN luminosity functions used.
(b) Relic \BHMF's obtained from the \protect\cite{ueda} AGN luminosity function 
but with different $\lambda$, \zo\ and $\delta$ with respect to panel (a).
For comparison, the thick line with no symbols indicates the relic \BHMF\
from \protect\citealt{ueda} plotted in (a).
As in the previous panels,
solid lines are the relic \BHMF's at $z=0$, while dotted lines are those at
$\zo$. }
\end{figure*}

\subsection{The Luminosity Function of Active Galactic Nuclei}

In order to ensure a consistent treatment, the AGN luminosity function \NAGN\
used in the continuity equation must describe the evolution of the {\it entire}
AGN population. The luminosity $L$ is the total luminosity radiated
from the accreting mass and is estimated with the bolometric corrections
described in the previous section.

The AGN luminosity functions available in the literature do not describe
the {\it entire} AGN population because they are the result of surveys performed
in "narrow" spectral bands, with given flux limits and selection criteria. For
example, the luminosity function of \cite{boyle} includes only quasars selected
for their blue color but it is now known there is a population of "red" quasars
which is missed.  The luminosity function of soft X-ray (0.5-2 keV) selected
AGNs of \cite{miyaji} misses most of the sources with significant X-ray
absorption ($N_H > \ten{22} \CM\2$); it includes mostly broad lined AGNs ($\sim
80\%$ of the total) but it is known that, at least locally, there is a dominant
population of obscured AGN's (e.g.~\citealt{maiolino95}).  Current hard
X-rays surveys (2-10 keV) are less sensitive to obscuration than optical and
soft X-ray ones, thus the very recent luminosity function by \cite{ueda} probes
by far the largest fraction of the whole AGN population. However it is
restricted only to Compton-thin AGNs ($N_H<\ten{24}\CM\2$), while it is known
that, at least locally, there is a significant fraction of Compton-thick
objects (e.g.~\citealt{risaliti}).  In summary, when using the luminosity
functions available in the literature, one should know that they describe part
of the AGN population, and thus account only for a fraction of the local \BHMF.

In this paper, we consider the AGN luminosity functions by
\citealt{boyle} (band $B$),
\citealt{miyaji} (0.5-2 keV) and \citealt{ueda} (2-10 keV).
Thus, the AGN luminosity function which will be used in the continuity equation
is given by
\begin{equation}
\phi(L)\, dL = \phi(L_x)\, dL_x
\end{equation}
where $x$ is either the $B$, the 0.5-2 keV or the 2-10 keV band and
$L=f_{bol,x}\, L_x$.

The AGN luminosity functions obtained with the bolometric corrections described
in the previous section are compared in Fig.~\ref{fig:agnlf}a at selected
redshifts. The luminosity functions by \cite{boyle} and \cite{miyaji} are in
rough agreement at the high $L$ end, meaning that they may be sampling the same
quasar population.  Indeed most of the objects in the \cite{miyaji} sample are
broad-lined AGN which, at high $L$, become the quasars observed by
\cite{boyle}.  The disagreement at low luminosities is because the
\citealt{boyle} luminosity function is extrapolated for $M_B>-23$ i.e.~for
$L<10^{12}\Lsun$.  In contrast the luminosity function by \cite{ueda} samples a
larger fraction of the AGN population at all luminosities.

In
Fig.~\ref{fig:agnlf}b we compare the differential comoving energy densities
computed in the luminosity ranges
$41.5 < \log (L_X/\ERG\S\1) < 48$ ($0.5-2\KEV$ and
$2-10\KEV$ bands) and $-28 < M_B < -21$ ($B$ band).  For the \cite{ueda}
luminosity function we also plot the differential comoving energy density for
objects with $L>10^{12}\Lsun$.  High luminosity objects provide $\sim 50\%$ of
the total energy density in the \cite{ueda} luminosity function and their
emission has a redshift distribution similar to that of the AGN's by
\cite{miyaji} and \cite{boyle}. Clearly, lower luminosity objects contribute
significantly at $z<1.5$ and this is an important result of the recent Chandra
and XMM surveys (e.g. \citealt{hasinger,fiore03,ueda}).
The total comoving energy densities in the
redshift range $z=0-3$
(i.e.~the integrals of the quantities plotted in the figure) are
4.2\xten{-16}, 6.4\xten{-16} and 1.5\xten{-15}\ERG\CM\3, 
for \citealt{boyle}, \citealt{miyaji}, and \citealt{ueda},
respectively. The high $L$ objects in the \cite{ueda} luminosity function 
provide 8\xten{-16}\ERG\CM\3, i.e.~$\sim 50\%$ of the total
energy density.

By applying Eq.~\ref{eq:chokshi}, the mass densities in BH's
can be written as
\begin{equation}
\rhoBH = \rho\,\left[\frac{1-\eff}{9\eff}\right] \rhoBHunits 
\end{equation}
where $\rho$ is 0.6 (from the \citealt{boyle} AGN LF), 0.9 (\citealt{miyaji}),
2.2 (\citealt{ueda}) and 1.2 (\citealt{ueda}, high $L$ objects).
The value derived from the \cite{ueda} luminosity function ($2.2\rhoBHunits$)
is already in marginal agreement with
$\rhoBH=4.5^{+1.8}_{-1.4}\rhoBHunits$, i.e.~the estimate of the local BH
density given in Sec.~\ref{sec:localbhmf}. 

\subsection{The Relics of Active Galactic Nuclei}
\label{sec:relics}

Given the AGN luminosity functions (\citealt{boyle,miyaji,ueda})
we can integrate the continuity equation assuming that $\lambda=1$
(AGN's emitting at the Eddington luminosity), $\eff=0.1$ and \zo=3.
Though we know that the above luminosity functions do not describe the whole
AGN population, initially we do not apply any correction for the
objects which have been missed.

The relic BH mass functions are shown in Fig.~\ref{fig:relics1}a.  As
expected, the AGN's traced by the \cite{ueda} luminosity function leave
more relics than those traced by the luminosity functions of \cite{boyle}
and \cite{miyaji}.  It is also notable that the relic \BHMF's at $z=0$
(solid lines) are $\sim 2$ order of magnitude larger than those at $\zo=3$
(dotted lines) meaning that most of the BH growth took place for $z<\zo=3$.  The
dashed lines represent the relic \BHMF\ obtained by considering only AGN's with
$L > 10^{12} \Lsun$. [In practice this was obtained by multiplying the AGN
luminosity functions by $\exp(-10^{12}/L)$.] The comparison between the dashed
and solid lines indicate that today high mass BH's ($\MBH>10^8\Msun$) grew
during quasar phases ($L>10^{12}\Lsun$). 

\begin{figure*}
\centering
\includegraphics[width=0.48\linewidth]{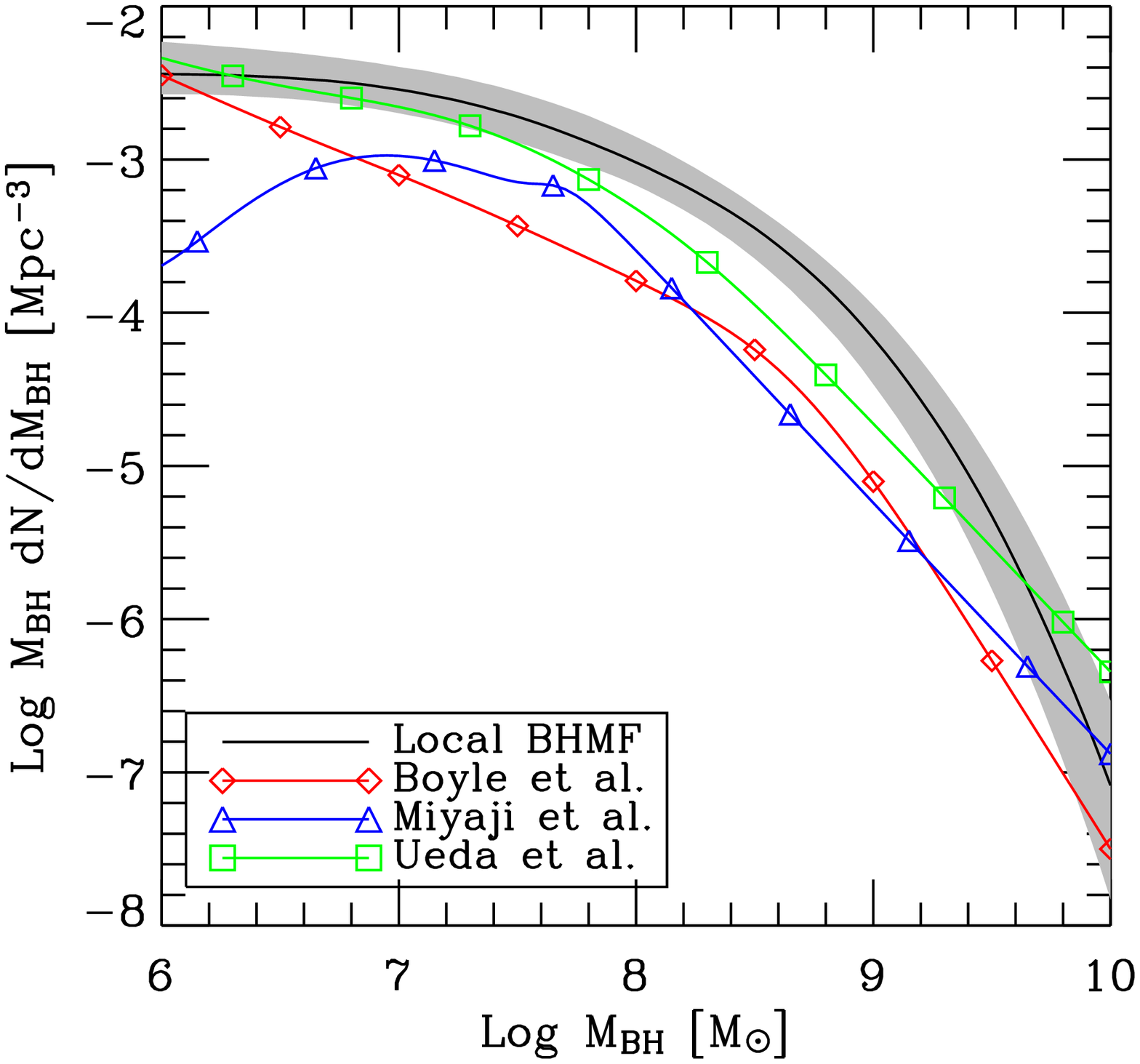}
\includegraphics[width=0.48\linewidth]{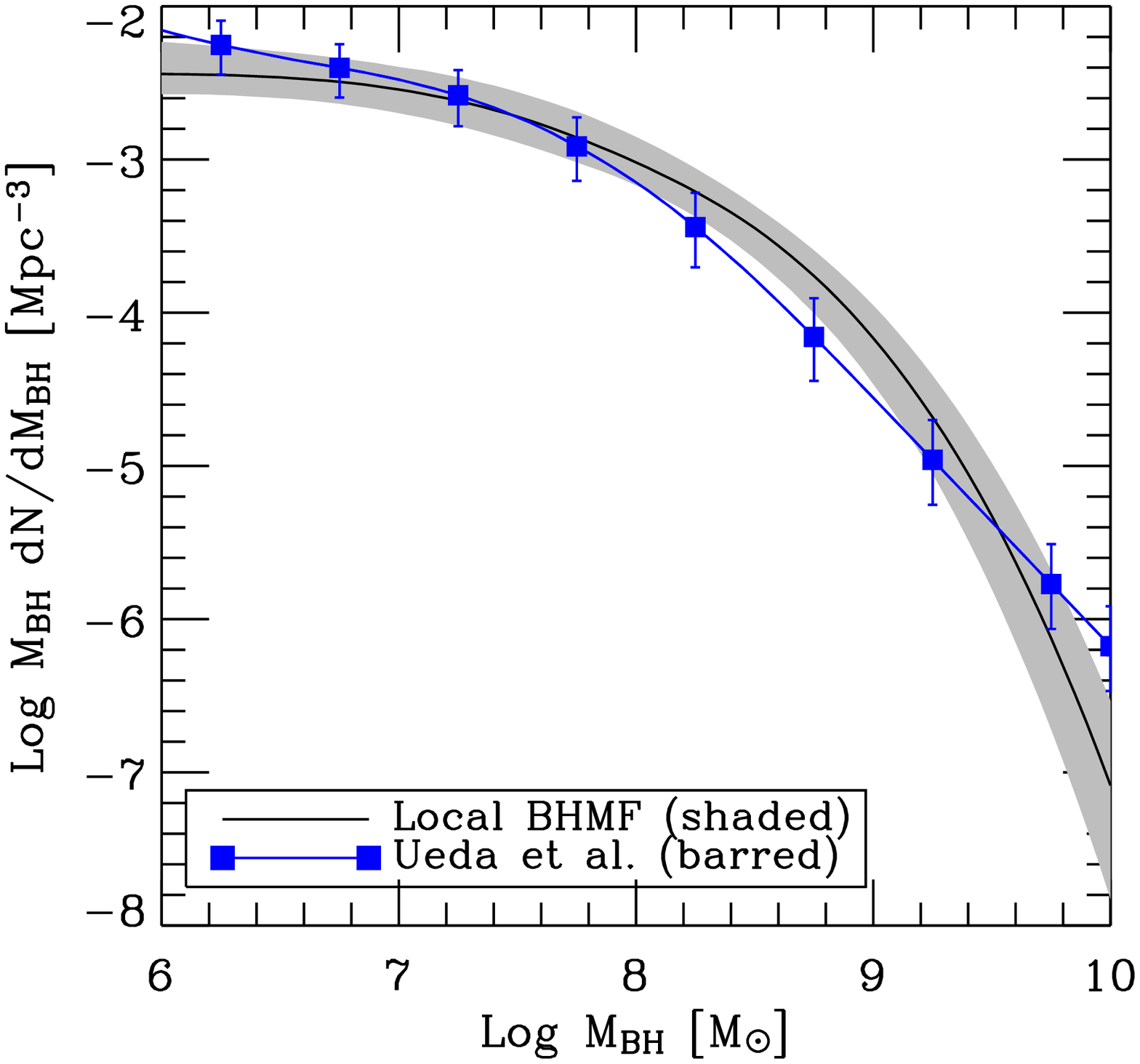}
\caption{\label{fig:relics2} (a) Local 
and relic \BHMF's. The shaded area represents the $\pm 1\sigma$
uncertainty on the logarithm of local \BHMF\ (see Fig.~\ref{fig:bhall}b).
(b) Relic \BHMF\ obtained using the \protect\cite{ueda}
hard X-ray luminosity function
and accounting for the missing Compton Thick AGN's (see text).
The errorbars 
are the 16 and 84\%\ percentile levels of 1000 realizations of the
relic \BHMF\ (see text).
The local \BHMF\ is plotted with the same notation as in the panel (a).
}
\end{figure*}
The \BHMF's in Fig.~\ref{fig:relics1}a were estimated assuming $\lambda=1$ and
$\eff=0.1$.  It can be seen from Eq.~\ref{eq:NBH} that the efficiency $\eff$
is a simple scaling factor and an increase in $\eff$ will decrease the level of
the relic \BHMF. However  a variation of the Eddington fraction $\lambda$ will
not have a simple scaling effect but it will produce a combination of scaling
and translation along the $M$ axis since $\lambda$ also enters the $L = \lambda
M c^2/t_E$ relation.  In Fig.~\ref{fig:relics1}b the same relic \BHMF\ derived
using \citealt{ueda} (thick line with no symbols) is compared with the \BHMF\
obtained assuming $\lambda=0.1$ (line with empty squares).  As previously,
solid and dotted lines indicate the \BHMF\ at $z=0$ and at $z=\zo$
respectively.  Decreasing $\lambda$ has the net effect of increasing the number
of \BH's at high masses ($M>10^{8.5}\Msun$) while decreasing that at lower
masses. In the same figure, we also show the \BHMF's obtained by assuming that
the starting redshift of integration is $\zo=5$ (line with empty triangles) and
$\zo=2.5$ (line with empty diamonds) and that the fraction of active BH's at
$z_0$ is $\delta=0.1$ instead of 1 (line with stars).  The relic \BHMF\ at $z=0$
is almost independent of the initial conditions.  Changing the starting
assumption (\zo, or $\delta[\zo]$) does not have any appreciable effect
provided that $\zo \ga 2.5$.  This is a consequence of the fact that the main
BH growth takes place at $z<3$ when there is more time available ($\sim 85$\%
of the age of the universe). At larger redshifts there is too little time for
the BH growth.

In summary, relic BH's grew mainly at low redshifts ($z<3$) and high mass
BH's were produced during quasar activity.  The initial conditions for the
continuity equation influence little the relic \BHMF\ at $z=0$.

\section{Local BH's and AGN Relics}
\label{sec:localrelics}

In this section we compare the local and relic \BHMF's.  At first
we will consider the relic \BHMF's derived in the previous section, without
trying to account for the AGN population missing in the adopted luminosity
functions.  After showing that, as expected, the \cite{ueda} luminosity
function encompasses the largest fraction of the AGN population, thus providing
a better match to the local \BHMF, we will focus on it and try to account for
the missing AGN's in a way which also satisfies the constraints imposed by the
X-ray background.

In Fig.~\ref{fig:relics2}a
we compare the local \BHMF\ derived in Sec.~\ref{sec:localbhmf} (Fig.~\ref{fig:bhall}b)
with the relic \BHMF's obtained adopting the \cite{boyle}, \cite{miyaji},
and \cite{ueda} AGN luminosity functions ($\eff=0.1$, $\lambda=1$)
without any correction for the missing AGN population.
As shown in the previous section these relic \BHMF's are insensitive
to the adopted initial conditions thus the parameters 
they depend on are only $\eff$ and $\lambda$.

The first result which can be evinced from the figure is that there is no
discrepancy at high masses.  All the relic \BHMF's are slightly smaller than
the local \BHMF. As explained previously, this is not significant because all
the adopted luminosity functions are missing part of the
AGN population. Conversely \cite{yu02} and \cite{ferrarese} using the
\cite{boyle} luminosity function found a significant 
discrepancy because 
their relic \BHMF\ at high masses is larger than the local one.
The reasons for the solution of this discrepancy are the following.
\begin{itemize}
\item We have taken into account the intrinsic dispersion of the
\MBHsig\ and \MBHlum\ correlations thus softening the high mass
decrease of the local \BHMF.
\item We have adopted zero points in the \MBHsig\ and \MBHlum\ correlations
which are a factor $\sim 1.6$ ($\sim 0.2$ in log) larger than those used by
\cite{yu02}. These zero points (and slopes) were derived from the analysis of
\cite{MH03} and their larger value is a consequence of the rejection of
galaxies with unreliable \BH\ masses.
\item We have adopted bolometric corrections which, on average, are $\sim 2/3$
of those adopted by previous authors. This is because we have not taken into
account the IR emission in the estimate of the bolometric luminosity. The IR
bump is produced
by reprocessed UV radiation and thus its use in the determination
of $L$ would result in overestimated accretion rates.
\end{itemize}

We now focus on the \cite{ueda} AGN luminosity function since
we have established that, as expected, it is the one which provides the
most complete description of the AGN population.
The aim is to assess whether the relic \BHMF\ can account for the whole
local \BHMF, i.e.~if local \BH's can be entirely explained as being
AGN relics.

The \cite{ueda} luminosity function only describes the population of
Compton-thin AGN's, i.e.~objects with absorbing column densities $\log N_H <
24$ [\CM\2]. To estimate the correction for the missing Compton-thick sources,
we follow \cite{ueda} and assume that there are as many AGN's in the 24-25 bin
of $\log N_H$ as in the 23-24 one, as indicated by the $N_H$ distribution of
\citealt{risaliti}. Thus, with the $N_H$ distribution estimated by \cite{ueda},
the luminosity function has to be multiplied by 1.3 in order to account for the
Compton-thick AGN's. In reality, the correcting factor is luminosity dependent
but, since it is limited to the $1.2-1.4$ range, we chose an average value for
simplicity.
With this contribution from
Compton-thick sources, \cite{ueda} are able to provide a reasonable fit of the
X-ray background spectrum. (We will return to this issue in the following
section.) While AGN's at $\log N_H>25$ are not important contributors of the
X-ray background, they do contribute to the relic \BHMF.  Thus, we assume
that for $\log N_H>25$ there are as many AGN's as in the 23-24 or 24-25 $\log
N_H$ bins. This assumption is also justified by the \cite{risaliti} $N_H$
distribution. The correcting factor becomes then 1.6.
Therefore, to compensate for the missing obscured AGN's we apply a
correcting factor of 1.6 independently of luminosity. In Fig.~\ref{fig:relics2}b
the corrected \cite{ueda} luminosity function is used to
determine the relic \BHMF.  We also show the uncertainties (16 and 84\%\
percentiles, i.e.~$\pm 1\sigma$ errors on the log of the relic \BHMF\ as previously) estimated
with the usual 1000
Montecarlo realizations of the relic \BHMF.
These were obtained by varying the number density of the
luminosity function ($\pm 30\%$ $1\sigma$
error on the number density to avoid
correlated errors on the other parameters),
the hard X-ray bolometric correction
($\pm 0.1$ $1\sigma$
error on $\log f_{bol,X}$), and the factor used to correct for the missing Compton-thick AGN's ($\pm 0.3$ $1\sigma$ error).

The relic and local \BHMF\ agree well within the uncertainties. 
Thus, adopting the best possible description of the whole AGN population,
the mass function of relic \BH's is in excellent agreement with the mass
function of local \BH's. Local \BH's are thus AGN relics and were mainly grown
during active phases in the life of the host galaxy.
This agreement has been obtained
with $\eff=0.1$ and $\lambda=1$ indicating that (i) efficiencies higher than
commonly adopted for AGN's are not required, and (ii) the main growth of BH's
occurs in phases during which the AGN is emitting close to
the Eddington limit.
In Section \ref{sec:efflambda} we will explore the locus in the $\lambda-\eff$
plane which is permitted by the comparison from the local and relic \BHMF's.
As discussed in Section \ref{sec:conteq}
we have neglected merging in our estimate of the
relic \BHMF. However, the good agreement between the local and relic \BHMF's
suggests that the merging process does not significantly affect the build-up of
the \BHMF, at least in the $z<3$ redshift range. 

\section{Constraints from the X-ray Background}
\label{sec:xrb}

From the X-ray background (XRB) light it is possible to estimate the expected
mass density of relic BH's (\citealt{salucci, fi99, erz02}) which can then be
compared with the mass density of local BH's.  From this comparison,
\cite{erz02} inferred that massive BH's must be rapidly rotating for the high
efficiency needed ($\eff>0.15$) to match the XRB and local BH mass densities.
We will show that, due to the redshift distribution inferred from the new X-ray
surveys, the match between the XRB and local BH mass densities can be obtained
without requiring large efficiencies ($\eff > 0.1$;
see also \citealt{fabian03,comastri03}).

The XRB provides another type of constraint. It has been shown (\citealt{setti,madau94,comastri95,gilli99}) that the XRB spectrum can be reproduced
by summing the spectra of the whole AGN population after suitable corrections
to take into account the absorption along the line of sight. The AGN population
and its redshift distribution is derived from AGN luminosity functions but, in
order to fit the XRB spectrum, one has to include a correction for the missing
(obscured)
AGN's. This correction is the same one which must be
adopted here to determine the relic
\BHMF\ from the whole AGN population.  Thus, the ratio $R_{12}$ between AGN's
included and missed in the adopted luminosity functions is a parameter which
determines both the relic \BHMF\ and the XRB spectrum (and the X-ray source
counts).  When rescaling the AGN luminosity function with the
$(1+R_{12})$ factor to match the local \BHMF, one should also verify that it is
also possible to match the XRB spectrum and source counts at the
same time. In practice, one can use the $R_{12}$ value required by the XRB synthesis models in order to fit the XRB spectrum and source counts.
It is beyond the
scope of this paper to produce a model synthesis of the XRB 
but we will show that with $R_{12}$ values found in
the literature we can match the local and relic \BHMF, and also
satisfy the constraints imposed by the X-ray background.

The local density in massive BH's expected from the observed X-ray background
(XRB) light can be estimated with the relation 
\begin{equation}
\rhoBH = \frac{1-\eff}{\eff c^2}\,(1+\langle z\rangle)\,U_T^\star
\end{equation}
where $U_T^\star$ is the total {\it observed} (as opposed to {\it comoving})
AGN energy density and
$\langle z\rangle$ is the average source redshift.
The factor $(1-\eff)$ is needed to
take into account the fact that not all the accreting mass falls into the \BH.
$U_T^\star$ can be estimated from the {\it observed} X-ray background light
as
\begin{equation}
U_T^\star = f_{bol,X}\times f_{obsc}\times \frac{4\pi}{c} I_{X}
\end{equation}
where $I_X$ is the total observed surface brightness of the X-ray background
(i.e.~the integral of the XRB spectrum), $f_{obsc}$ is the correction
to take into account source obscuration in the X-rays
(i.e.~$f_{obsc}\times I_{X}$ would be the total XRB surface brightness if AGN's were not
obscured)
and $f_{bol,X}$ is the X-ray bolometric correction
(see for more details \citealt{fi99, salucci, erz02}).
\cite{erz02} estimate $U_T^\star\sim (1.5 - 3.4)\xten{-15}\ERG\CM\3$.

We first verify that the above formula is consistent with the
scheme followed in this paper, and then establish how
$\langle z\rangle$ must be computed.
The observed background surface brightness at energy $E$ is
\begin{eqnarray}
\lefteqn{I(E) = \frac{1}{4\pi} \times} \nonumber \\
& & \int_0^{z_0} dz\, \frac{(1+z)}{4\pi D_L^2}
\frac{dV}{dz} \int_{\tilde L_1}^{\tilde L_2}
f[E(1+z)] \tilde L\phi(\tilde L,z)d\tilde L
\end{eqnarray}
where $D_L$ is the luminosity distance, $V$ is the comoving volume,
$\tilde L$ is the source luminosity in the energy band $x$ and
$\phi$ is the luminosity function in the same band.
$f[E]$ is the source spectrum at energy $E$ normalized to have unit
luminosity in the band $x$. 
Integrating on $E$ to find the total
surface brightness in band $x$ one finds
\begin{equation}
I_x = \frac{c}{4\pi} \int_0^{z_\mathrm{max}} dz \frac{1}{(1+z)} \frac{dt}{dz}
\int_{\tilde L_1}^{\tilde L_2} \tilde L \phi(\tilde L,z) d\tilde L
\end{equation}
Applying the bolometric correction ($L=f_{bol,x} \tilde L$ 
with $\phi(\tilde L,z) d\tilde L$ = $\phi(L,z) dL$),
the obscuration correction ($f_{obsc}$) and comparing
with Eq.~\ref{eq:Ecomov}, one finds that 
\begin{equation}
(1+\langle z \rangle) \frac{4\pi I_T}{c} =
(1+\langle z \rangle) U_T^\star = U_T
\end{equation}
where $I_T=f_{bol,x} f_{obsc} I_x$ and
$U_T$ is the total {\it comoving} (as opposed to {\it observed})
energy density.
The average redshift $\langle z \rangle$ is then
\begin{equation}\label{eq:zav}
(1+\langle z \rangle) = \frac{ \int_0^{z_\mathrm{max}} dz \frac{dt}{dz}
\int_{L_1}^{L_2} L \phi(L,z) dL } 
{ \int_0^{z_\mathrm{max}} dz \frac{1}{(1+z)} \frac{dt}{dz}
\int_{L_1}^{L_2} L \phi(L,z) dL }
\end{equation}

Using the luminosity function by \cite{ueda} one finds
$\langle z\rangle=1.1$, 
lower than the value $\langle z\rangle\sim 2$
assumed by previous authors. For comparison, the high
luminosity objects ($L>10^{12}\Lsun$) in the \cite{ueda}
luminosity function have $\langle z\rangle=1.3$ while
the \cite{miyaji} and \cite{boyle} have $\langle z\rangle=1.4$
and $\langle z\rangle=1.5$, respectively.
With the $U_T^\star$ estimate by \cite{erz02} we get
\begin{equation}
\rhoBH=(4.7-10.6) \left[\frac{(1-\eff)}{9\eff}\right]\rhoBHunits
\end{equation}
which is perfectly consistent with the estimate from the local BHMF
$\rhoBH=(3.2-6.5)\rhoBHunits$ without requiring efficiencies
larger than the `canonical' value \eff=0.1. This agreement has also
been remarked by \citealt{fabian03} and \citealt{comastri03}.
The critical point is clearly the value
of $\langle z\rangle$, the average redshift of X-ray sources emitting the XRB,
which has been significantly reduced by Chandra and XMM surveys
(e.g.~\citealt{hasinger,fiore03,steffen,ueda}).
The minimum value of \eff, allowed for a
consistency between the two estimates of \rhoBH, is $\eff=0.07$
just slightly larger than 
the non-rotating BH case.  Conversely, the maximum allowed efficiency is
$\eff = 0.27$, substantially below the maximally rotating Kerr BH case
(see Sec.~\ref{sec:efflambda}).
It is intriguing to find
that efficiencies smaller than those expected from non-rotating BH's or larger that those expected from maximally rotating Kerr BH's
are excluded.

We have thus verified that the expected density of BH remnants inferred from
the XRB is consistent with the local one without requiring efficiencies larger
than the canonically adopted value $\eff=0.1$.
We now verify if, with the
obscured/unobscured ratios $R_{12}$ adopted in XRB synthesis models, it is
possible to reproduce also the local \BHMF.
For the scope of this paper it suffices to notice that the
\cite{ueda} luminosity function,
with the correction for the missing Compton-thick AGN's that we also 
adopt, is used by the same authors to successfully reproduce 
the XRB spectrum and source counts.  Thus
the agreement of
the local and the relic \BHMF\ in Fig.~\ref{fig:relics2}b is obtained by also
meeting the constraints from the XRB which, in practice,
provide an estimate of the number of AGN's {\it
missed} by the luminosity function.
The same comparison could also be done using the \cite{miyaji} luminosity
function combined with the background model of \cite{gilli01}.  However, though
that model is successful in reproducing the XRB spectrum and source counts,
recent Chandra and XMM surveys have shown that the model redshift distribution 
is not correct (e.g.~\citealt{hasinger}).

In summary, our analysis shows that it is possible to meet the XRB constraints
both in terms of \rhoBH\ and of the local \BHMF.  

\section{Accretion Efficiency and Eddington Ratio}
\label{sec:efflambda}

The relic \BHMF\ derived from the \cite{ueda} AGN luminosity function,
corrected for the Compton-thick AGN's, provides a good match to the local
\BHMF\ and also satisfies the XRB constraints.  The match is obtained for
$\eff=0.1$ and $\lambda=1$. Here we investigate the locus in the $\lambda-\eff$
plane where an acceptable match of the local and relic \BHMF's can be found,
i.e.~we determine the acceptable $\lambda$ and \eff\ values.

To quantify the comparison between the local and relic \BHMF's we
recall that all the realizations of the \BHMF's have a log-normal
distribution at given $M$ and we
consider the following expression:
\begin{equation}\label{eq:kav}
k^2 = \int \frac{\left[\log N_L(M)-\log N_R(M) \right]^2 } 
{ \sigma_L(M)^2+\sigma_R(M)^2 }\, \mathrm{d}\log M {\bigg /} \int \mathrm{d}\log M 
\end{equation}
where $N_L(M)$ and $N_R(M)$ are the local and relic \BHMF's
with their $1\sigma$ uncertainties, $\sigma_L(M)$ and $\sigma_R(M)$. 
The integration is performed in the $\log(M/\Msun)=6-10$ range.
$k^2$ is the average square deviation
between the logarithms of the two \BHMF's measured in units of the
total standard deviation. If, for instance, $\log N_L(M) = \log N_R(M) +
n \sqrt{\sigma_L(M)^2+\sigma_R(M)^2}$, then $k^2=n^2$, i.e.~the
two functions differ, on average, by $n$ times the total standard
deviation.

\begin{figure}
\centering
\includegraphics[width=0.98\linewidth]{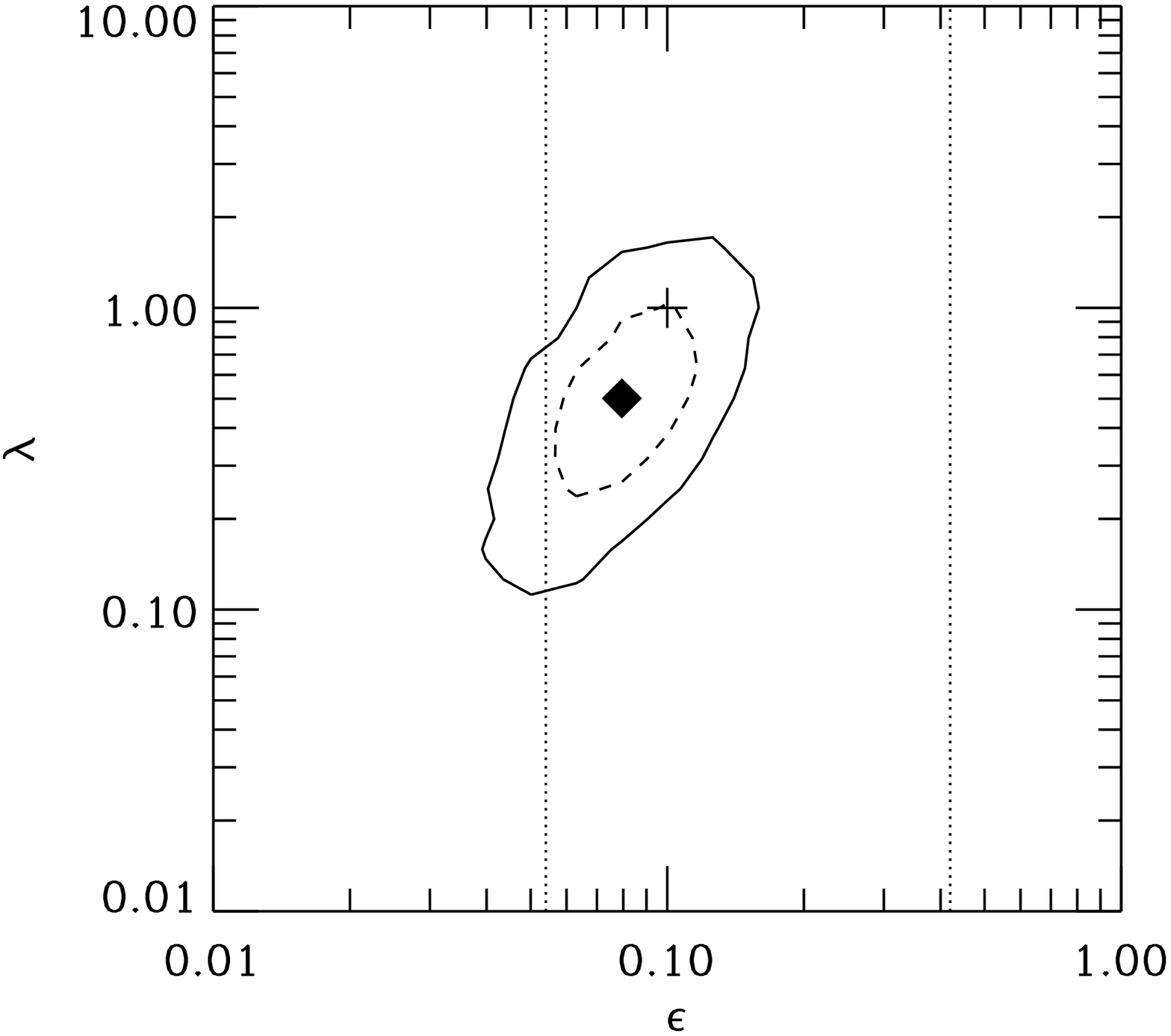}
\caption{\label{fig:xrb}
\label{fig:efflambda} 
Locus where 
accretion efficiency \eff\ and  Eddington ratio $\lambda$
provide the best match between the local
and relic \BHMF's. The solid and dashed lines limit the
regions where the logarithms of the local
and relic \BHMF\ differ, on average, by less than 1 and 0.7$\sigma$
(see text). 
The filled diamond marks the $k^2$ minimum and corresponds to
$\eff=0.08$ and $\lambda=0.5$. The cross indicates $\eff=0.1$ and $\lambda=1$.
The dashed lines are theoretical efficiencies
for non-rotating ($\eff=0.054$) and maximally
rotating \BH's ($\eff=0.42$).
}
\end{figure}

The $\eff$ and $\lambda$
values that corresponds to the minimum $k^2$ ($k^2_\mathrm{min}=0.4^2$)
are marked by the filled square and are $\eff=0.08$ and
$\lambda=0.5$.
To have an acceptable match between the local and relic \BHMF\ we require that
$k^2\le 1$ and this constraint identifies 
the region limited by the solid line in Fig.~\ref{fig:efflambda}.
For comparison, the dashed line
limits the region where $k^2<0.7^2$, which corresponds to 
the average square deviation in the "canonical" case, $\eff=0.1$
and $\lambda=1$, which is marked by the cross.
The allowed efficiencies include the 
non-rotating Schwarzschild \BH\
($\eff=0.054$, \citealt{schwarzschild,shapiro83})
and are well below
the maximally
rotating Kerr \BH\ ($\eff=0.42$, \citealt{kerr63,shapiro83}).
The best agreement between the local and relic \BHMF's is obtained
for efficiencies larger than that of the non-rotating \BH\ case
suggesting that, on average, \BH's should be
rotating.  \cite{hughes} found that \BH's are typically spun down by mergers
and this would limit the importance of mergers in the growth of \BH's, in
agreement with our assumption in the continuity equation.  

The allowed Eddington ratios, $\lambda=L/\LEDD$, are in the range
$0.1<\lambda<1.7$ indicating that \BH\ growth takes place during luminous
accretion phases close to the Eddington limit.  \cite{mclure03}, using a large
sample of SDSS quasars, have recently estimated
that the average $\lambda$ varies from 0.1 at
$z\sim 0.2$ to 0.4 at $z\sim 2$. Accounting for the different bolometric
corrections (they used $f_{bol,B}=9.8$, $\sim 1.5$ times larger than the value
adopted by us), $\lambda$ varies from 0.15 at $z\sim 0.2$ to 0.6 at $z\sim 2$.
These values are in excellent agreement with the constraints posed on $\lambda$
in Fig.~\ref{fig:efflambda}.

The results in Fig.~\ref{fig:efflambda} do not imply that accreting BH's
cannot have $\eff$ and $\lambda$ values outside the region of the
best match between the local and relic \BHMF's. Indeed, those
limits are only for average efficiencies and Eddington ratios 
during phases in which Black Holes are significantly grown.

We conclude by noting
that $\eff$ represents only the radiative efficiency.  If there is
significant release of mechanical energy, the true efficiency might be higher.
For instance, in M87 the kinetic energy carried away by the jet is much larger
than the radiated one (\citealt{owen}) and, in general, the jets of radio loud
AGN's can carry away up to half of the total power in 
kinetic energy (\citealt{rawlings91,celotti97,tavecchio00}).
The release of mechanical and radiative energy is important for
the feedback on the galaxy which is thought to be one
of the causes behind the \MBHsig\ and \MBHlum\
correlations (e.g.~\citealt{silk98,blandford99,begelman,granato04}).
Taking into account mechanical energy, the
expression $\eff/(1-\eff)$ should be transformed to $\eff_R/(1-\eff_R-\eff_M)$,
where $\eff_R$ and $\eff_M$ are the radiative and mechanical efficiencies.
Then, similarly, to what has been done in section, one could place
constraints on both $\eff_R$ and $\eff_M$ but this is beyond the
scope of this paper.
\begin{figure*}
\centering
\includegraphics[width=0.48\linewidth]{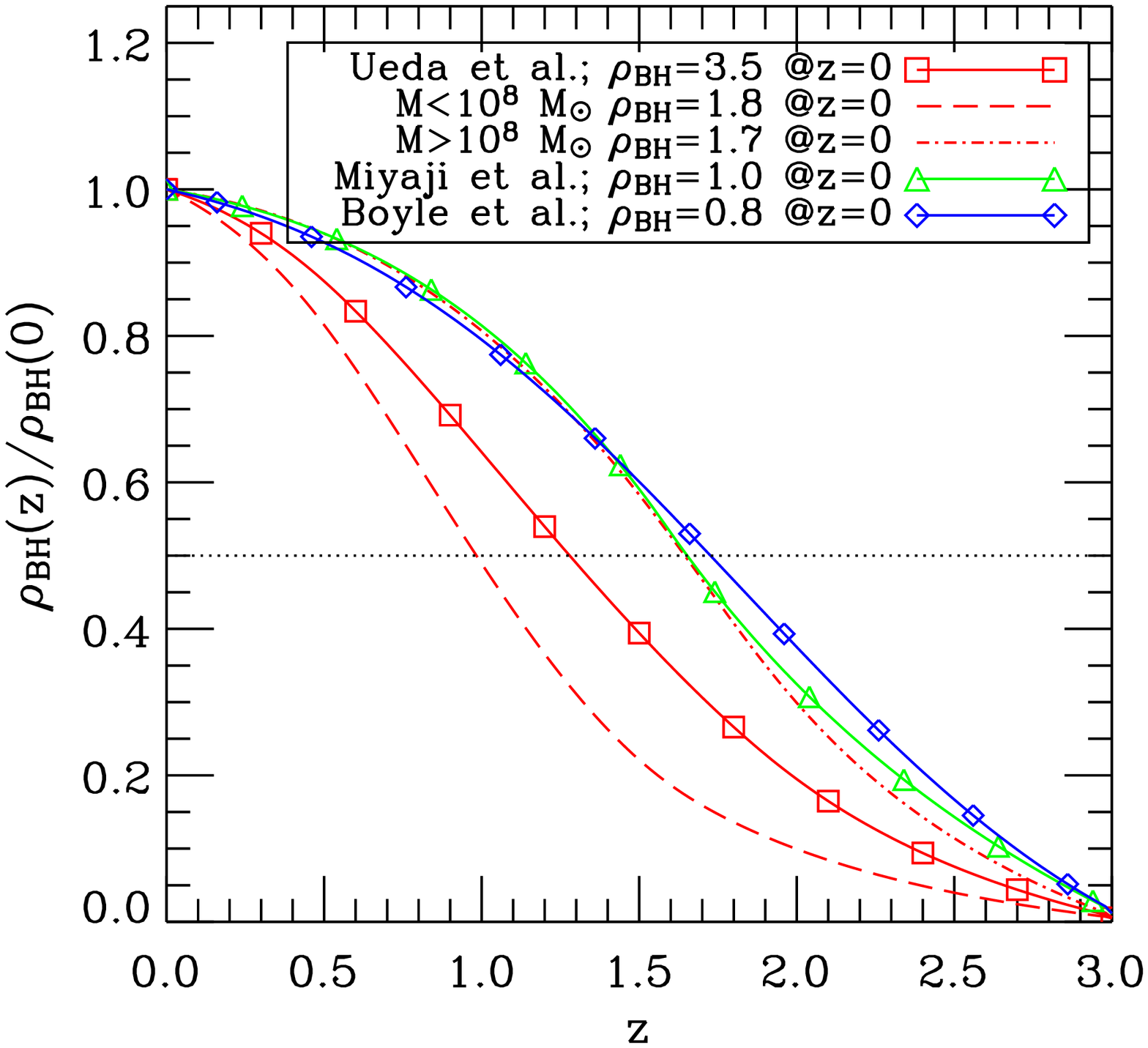}
\includegraphics[width=0.48\linewidth]{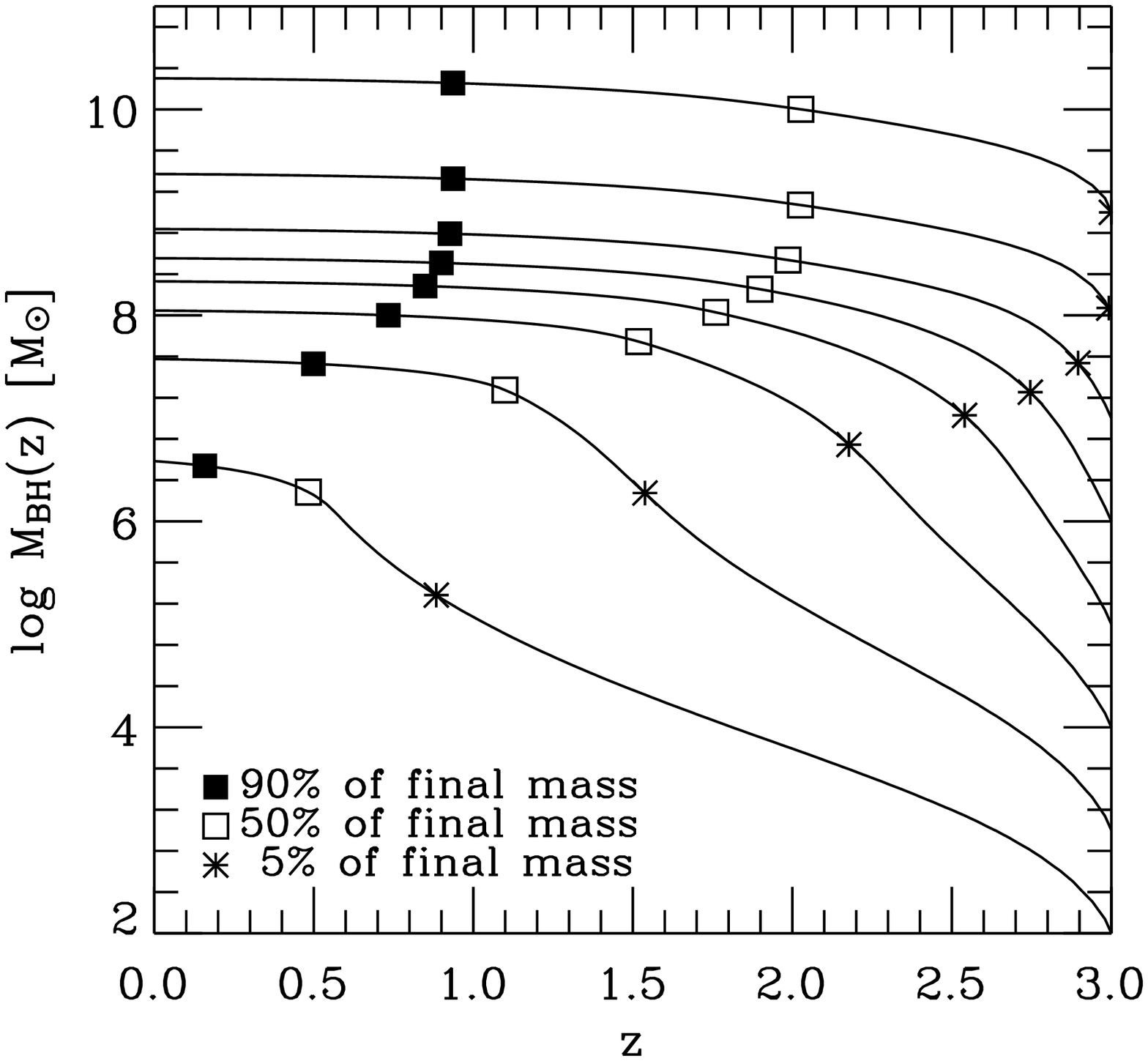}
\caption{ \label{fig:bhgrowth}
(a) Growth history of total BH mass density
(with $\eff=0.1$ and $\lambda=1$). Different symbols
indicate the use of different AGN luminosity functions. The dashed and dot-dashed lines indicate the growth history of \rhoBH\ in low
($M<\ten{8}\Msun$) and high mass \BH's, computed using the
\protect\cite{ueda} luminosity function. The densities are normalized to the
final value at $z=0$ which is reported in the upper part
of the plot. The units for \rhoBH\ are $\ten{5}\Msun\MPC\3$.
(b) Average growth history of \BH's
with given starting mass at \zo=3\ computed using the
\protect\cite{ueda} luminosity function and $\eff=0.1$, $\lambda=1$.
The symbols, filled squares, empty squares , and stars, indicate
when the \BH\ reaches 90\%, 50\%, and 5\%\ of its final mass,
respectively.
}
\end{figure*}

\section{Growth and Accretion History of Massive Black Holes}
\label{sec:hist}

Having established that the obscuration corrected AGN luminosity function by
\cite{ueda}
with $\eff=0.1$ and $\lambda=1$ provides a relic \BHMF\
which is fully consistent with the local \BHMF\ and the X-ray background,
we can analyze the growth history of massive \BH's and, in the
next section, the average lifetime of their active phases.

The redshift dependence of the total density in massive
BH's is given by (Eqs. \ref{eq:chokshi} and \ref{eq:Ecomov})
\begin{equation}
\rhoBH(z^\prime) =  \frac{1-\eff}{\eff c^2}\,
\int_{z^\prime}^{\zo} dz\,\frac{dt}{dz}\int_{L_1}^{L_2} L \phi(L,z)d\log L
\end{equation}
and is plotted in Fig.~\ref{fig:bhgrowth}a.
The relics of the AGN's traced with the luminosity functions
by \cite{boyle} and \cite{miyaji} reach 50\%\ of the $z=0$
mass density around $z\sim 1.7$, while the AGN's traced by the \cite{ueda}
luminosity function do so at $z\sim 1.4$. This is a consequence of the
larger number of low luminosity AGN's which are present at $z\sim 1$
(see also Fig.~\ref{fig:agnlf}b). Indeed, when separating the
contributions from low ($\MBH<\ten{8}\Msun$) and high mass \BH's 
which contribute 50\%\ each of the $z=0$ \BH\ density, it is clear that
low mass \BH's grow later than high mass \BH's. These, like
in the case of the AGN's traced by \cite{boyle} and \cite{miyaji},
reach 50\%\ of their final mass at $z\sim 1.7$.

This issue can be further investigated by computing the
average growth  history of a \BH\ with given starting, or final, mass.
The growth history of a \BH\ with given starting mass ${\MBH}_0$
at \zo\ can be estimated as follows.
The average accretion rate at $z$ (or $t$) is given from Eq. 15 as:
\begin{equation}\label{eq:mdot2}
\langle \Mdot(M,t) \rangle = \frac{1}{t_E\, \ln 10}
\frac{(1-\eff)\,\lambda}
	{\eff\,N(M,t)}
	\left[\phi(L,t)\right]_{L=\lambda\frac{M c^2}{t_E}}
\end{equation}
thus one can solve the following differential equation
to obtain the 'average' growth history of \BH's:
\begin{equation}\label{eq:mbhgrowth}
dM = \langle \Mdot(M,t) \rangle \frac{dt}{dz} \, dz
\end{equation}
In Fig.~\ref{fig:bhgrowth}b we plot the average growth history of \BH's with
different masses at $\zo=3$.  A supermassive BH like that of M87 or Cygnus A
($\MBH\sim 3\xten{9}\Msun$; \citealt{marconi97,tadhunter}) was already quite
massive ($\sim \ten{8}\Msun$) at \zo\ while a smaller BH like that of Centaurus
A (\citealt{marconi01}) was less massive, around $\MBH=\ten{6}\Msun$.  A
supermassive BH with $\MBH\sim \ten{9}\Msun$ at $\zo=3$ should now be over
$\ten{10}\Msun$. Indeed, the existence of very massive \BH's at high redshifts
is suggested by the detection of very luminous quasars and, in particular,
those detected at $z\sim6$ by the SDSS survey (e.g.~\citealt{fan03}). For
instance, \MBH\ for the farther quasar known ($z\sim6$) is estimated as
$\MBH=3\xten{9}\Msun$ (\citealt{willott}), and one would expect its local
counterpart to be more massive than $\ten{10}\Msun$. However, these quasars have
not been detected yet and it is not clear if this is because these
hypermassive \BH's are very rare
or they simply do not exist. In the latter
case there should be a physical reason which prevents a \BH\ from growing
beyond \ten{10}\Msun, possibly the feedback on the host galaxy mentioned in the
previous sections (see also \citealt{netzer03}).

From Fig.~\ref{fig:bhgrowth}b we can also infer a confirmation of what already
found in Fig.~\ref{fig:bhgrowth}a, namely that more massive \BH's grow
earlier.  The symbols in the figure, filled squares, empty squares, and stars,
mark the points when a \BH\ reaches 90\%, 50\%, and 5\%\ of its $z=0$ mass,
respectively.  It is clear that for $z<\zo=3$ all \BH's gain more than 95\%\ of
their final mass but \BH's which are now more massive than $\ten{8}\Msun$ had
already gained 50\%\ of their $z=0$ mass at $z\sim 2$.
Conversely, \BH's which have now masses
around $\MBH\sim\ten{7}\Msun$ grew very recently, at $z<1$.  Again, this is a
consequence of the luminosity function by \cite{ueda} in which the distribution
of lower luminosity AGN's peaks at $z\sim 1$.  Thus, the luminosity function of
\cite{ueda} points toward an anti-hierarchical growth of \BH's in the sense
that the largest \BH's were formed earlier.

If the correlations between \MBH\
and host galaxy properties were valid at higher redshifts (as is suggested by
the results of \citealt{shields03}) this would immediately imply that also the
most massive galaxies should form earlier, in contrast with the predictions of
current semi-analytic models of galaxy formation (see the introduction of
\cite{granato04} for more details and references). The detection of high mass
galaxies in sub-mm surveys is indeed more consistent with the `monolithic'
scenario in which massive ellipticals form at relatively high redshifts
(e.g.~\citealt{genzel03,granato04} and references therein).
\begin{figure}
\centering
\includegraphics[width=0.98\linewidth]{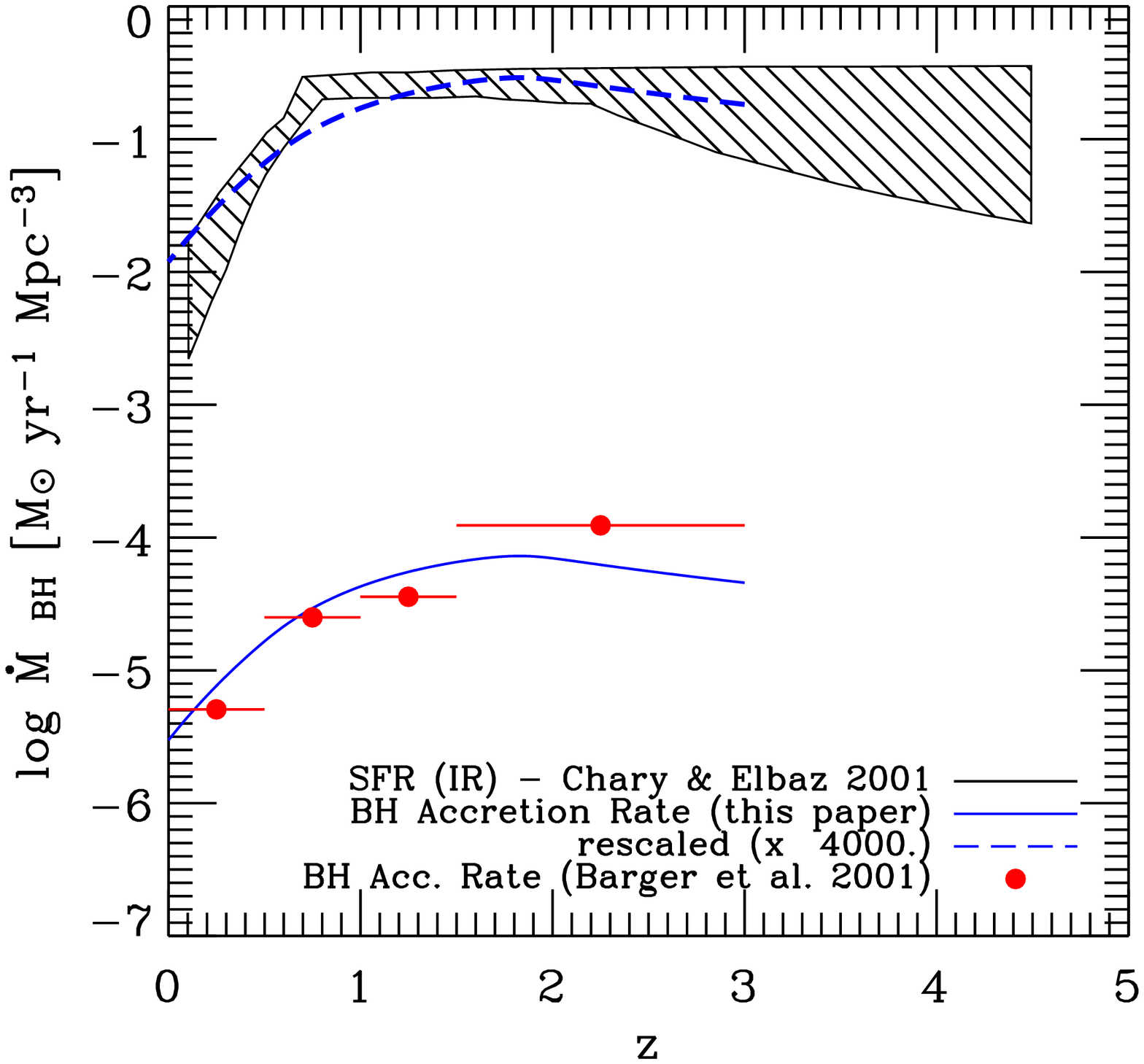}
\caption{ \label{fig:bhgrowth2}
Cosmic \BH\ accretion history compared with the estimate by
\protect\cite{barger01}
and with the cosmic star formation rate by \protect\cite{chary01}.}
\end{figure}

The BH accretion history (i.e.~the total accretion rate at given
$z$ per unit comoving volume)
can be estimated using Eq.~\ref{eq:mdot} as
\begin{equation}
\Mdot_{AGN}(z) =  \int_{M_1}^{M_2}N[M,t(z)] \langle \Mdot[M,t(z)] \rangle dM
\end{equation}
In Fig.~\ref{fig:bhgrowth2} we plot the cosmic accretion history
of \BH's (computed using the \cite{ueda} luminosity function,
$\eff=0.1$ and $\lambda=1$) and we compare
it with the estimate by \cite{barger01}
and with the cosmic star formation rate by \cite{chary01}.

Our estimate of the cosmic accretion rate onto \BH's agrees well with that
by \cite{barger01}, likely because most of the high redshift AGN's
used by \cite{ueda} come from the \cite{barger01} sample. Differently from us,
\cite{barger01} estimate the bolometric AGN luminosities by integrating the
observed spectral energy distributions. Thus the agreement with our analysis
should be viewed as a consistency check on the bolometric corrections and on
the corrections for obscured AGN's that we applied.

The cosmic accretion history has a similar redshift dependence as the
cosmic star formation rate, which we report here in the form
estimated by \cite{chary01}. To aid the eye, the dashed line
represents the estimate of the cosmic accretion history rescaled by 4000.
The comparison suggests that indeed the two rates have a similar redshift dependence and justifies the assumption that the accretion onto a BH
is proportional to the star formation rate, at least at a cosmic
level. This fundamental assumption is made in several models
of coeval evolution of \BH\ and galaxy and, together, with the 
feedback from the AGN explains the observed correlations \MBHsig\
and \MBHlum\ (e.g.~\citealt{granato04,haiman03}).

\section{The lifetime of active BH's}
\label{sec:time}

We now estimate the average lifetime of active BH's with the formalism used in
this paper and the corrected AGN luminosity function by \cite{ueda}.

\begin{figure*}
\centering
\includegraphics[width=0.48\linewidth]{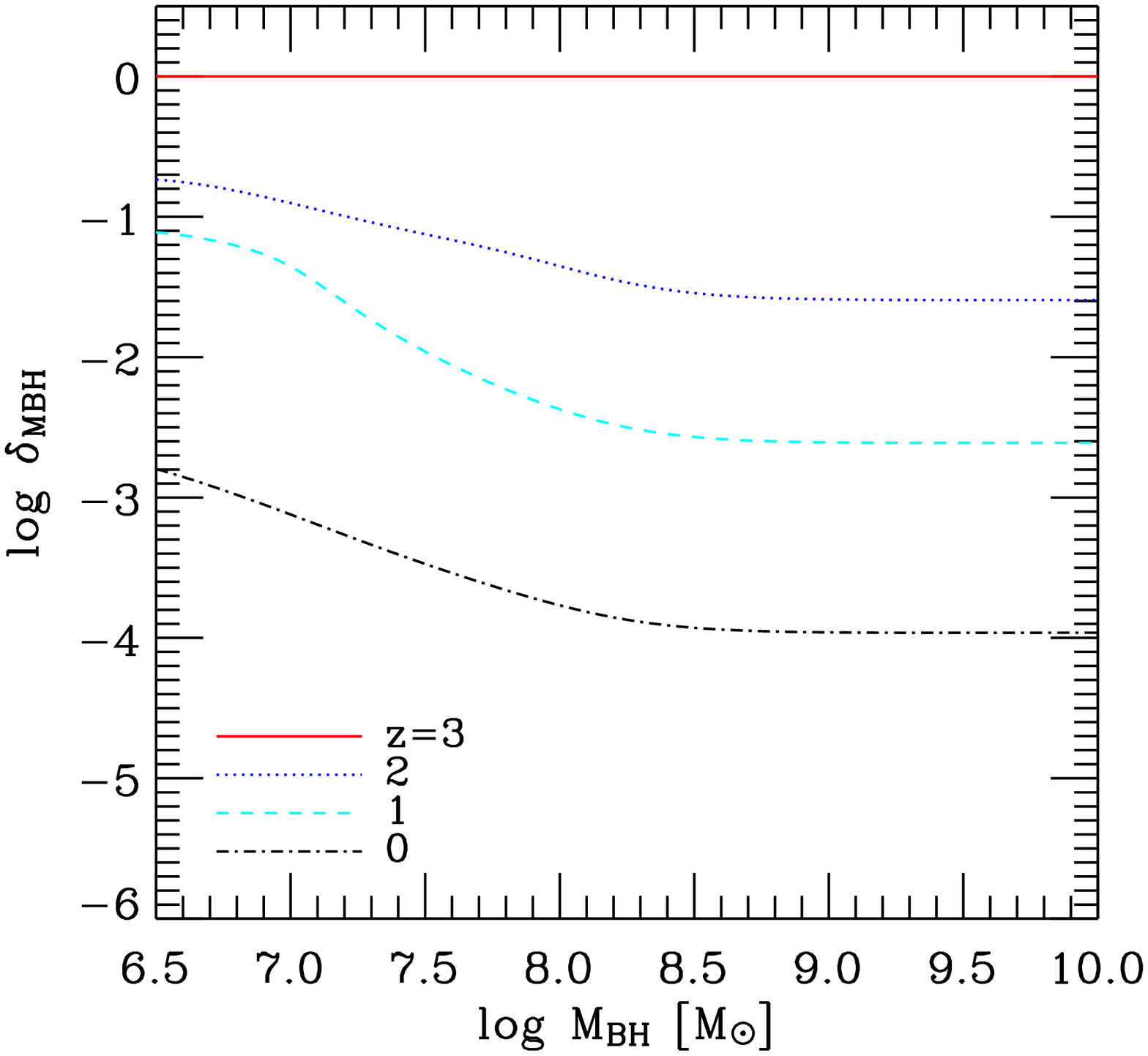}
\includegraphics[width=0.48\linewidth]{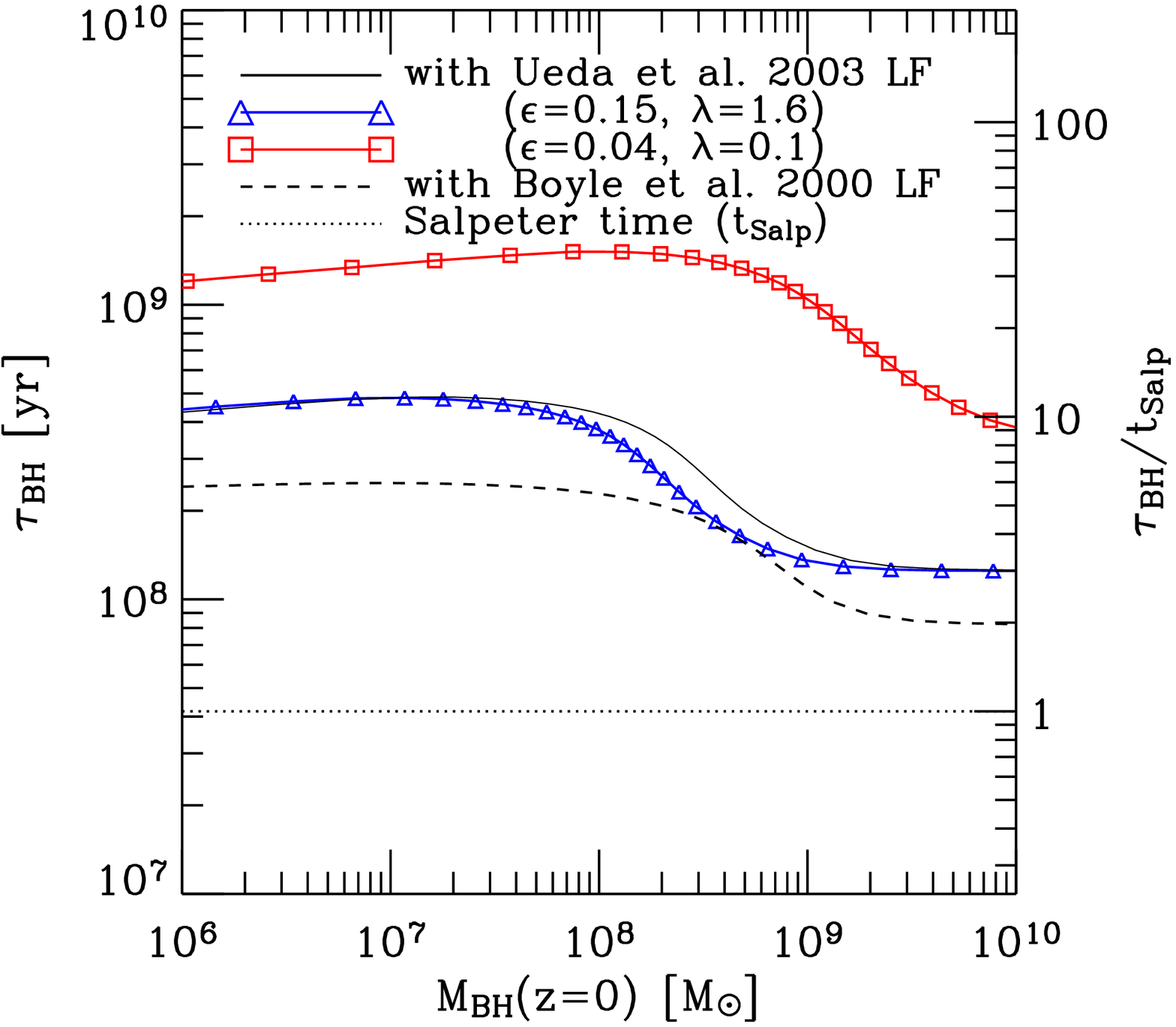}
\caption{\label{fig:bhduty}
(a) BH duty cycle $\delta(M,z)$ at given redshifts computed with the
\protect\cite{ueda} luminosity function and $\eff=0.1$, $\lambda=1$.
(b) Average mean lifetime of active \BH's (AGN's)
as a function of the relic \BH\ mass at $z=0$
computed with $\eff=0.1$, $\lambda=1$, the
\protect\cite{ueda} (solid line) and \protect\cite{boyle} (dashed line)
luminosity function. The dotted line
is the corresponding Salpeter time. The scale of the $y$ axis on the right
is the average lifetime in units of the Salpeter time.
The lines with the empty squares and triangles are the average
lifetimes computed from the \protect\cite{ueda} luminosity function with
$\eff=0.04$, $\lambda=0.1$, and $\eff=0.15$, $\lambda=1.6$, respectively.
}
\end{figure*}

In Fig.~\ref{fig:bhduty}a we plot the \BH\ duty cycle $\delta(M,t)$ at
selected redshifts derived from Eq. \ref{eq:phiagn} and computed using the
\cite{ueda} luminosity function, $\eff=0.1$ and $\lambda=1$.  From Sec.~\ref{sec:conteq},
the duty cycle $\delta(M,t)$ is the fraction of \BH's with mass $M$ active at
time $t$ or redshift $z$.  According to the definitions used in this paper, we
consider a \BH\ active if it is emitting at the
adopted fraction $\lambda$ of the
Eddington luminosity. Hence, objects
which are
usually classified as `active' but which are emitting well below their
Eddington limit (e.g.~M87 or Centaurus A),
should not be counted among the active BH's whose fraction is given
by the duty cycle.
\BH's more massive than $\sim\ten{9}\Msun$ are very rarely
active in the local universe (only 1 out of 10000) while they become more
numerous at higher redshifts (by a factor $\sim100$ at $z=2$).
Conversely, lower mass \BH's are
usually a factor 10 more numerous. A unit duty cycle at $z=3$ is the initial
condition assumed for the solution of the continuity equation (see
Secs.~\ref{sec:conteq} and \ref{sec:relics}).  The values of the
duty cycle we obtain at $z\sim 0-1$ are
consistent with the average values of $3-6\xten{-3}$ estimated by
\cite{haiman03}. 

The average duration of the accretion process, i.e.~the mean
lifetime of an AGN which has left a relic of mass $M$, is estimated by 
first solving Eq.~\ref{eq:mbhgrowth} to obtain $M(z, M_0)$, the growth history of a BH with
mass $M_0$ at \zo.  Then the 'active' time is simply given by
\begin{equation}
\tau(M) = \int_0^{\zo} \delta[ M(z, M_0), z ] \frac{dt}{dz} dz
\end{equation}
In Fig.~\ref{fig:bhduty}b we plot the average mean lifetime of AGN's (solid
line) as a function of the relic BH mass at $z=0$ computed using the
\cite{ueda} AGN luminosity function and $\eff=0.1$, $\lambda=1$.
The average lifetime of AGN's which leave a relic BH mass
of $> \ten{9}\Msun$ is of the order of 1.5\xten{8}\YR, while
for smaller relic masses ($<\ten{8}\Msun$) longer active phases
are needed ($\tau_{BH}\sim 4.5\xten{8}$\YR).
Considering two limiting cases 
from Fig.~\ref{fig:xrb}, $(\eff, \lambda)=(0.04,0.1)$, and $(0.15,1.6)$, 
the average lifetimes can increase up to $\ten{9}\YR$.
We remark that these
numbers are the average lifetimes for $z<\zo$ with \zo=3.

The derived lifetimes are consistent 
with the main result from this paper 
that most of the \BH\ masses are assembled via mass accretion.
Indeed, if a \BH\ grows by accreting matter with efficiency \eff\ 
and emitting at a fraction $\lambda$
of its Eddington luminosity, its e-folding time is given
by $t_s$, the Salpeter time (\citealt{salpeter}), 
\begin{equation}\label{eq:salptime}
t_{Salpeter} =
\frac{\eff\,t_E}{(1-\eff)\,\lambda} = 4.2\xten{7}\YR
\left[\frac{(1-\eff)}{9\eff}\right]^{-1} \lambda^{-1}
\end{equation}
Thus, the initial \BH\ mass has been e-folded $\sim 3$ times from $\zo=3$ for
$\MBH>\ten{9}\Msun$ and more than 10 times for $\MBH<\ten{8}\Msun$.
The apparently long lifetimes are thus the natural consequence of the
fact that 
to grow a \BH\ from small ($\sim 1-10\Msun$) or intermediate
mass seeds ($100-1000\Msun$, e.g.~\citealt{schneider02}) 
several e-folding times must pass easily implying $\tau_{BH}>\ten{8}\YR$.
For instance, to grow a \BH\ from \ten{3}\Msun\ 
to \ten{9}\Msun\ with $\lambda=1$
and $\eff=0.1$, one would need $\sim 14 t_s$ i.e.~$\sim 6\xten{8}\YR$
(see also \citealt{haiman}).
Indeed, in the model by \cite{granato04} of coeval evolution of
\BH\ and galaxy, the time needed to grow most of the \BH\ mass
is $\approx 3\xten{8}\YR$ (note however that they use
$\eff=0.15$, $\lambda=3$ with which the Salpeter time is roughly halved 
with respect to our paper).

The reason why the growth of smaller \BH's ($<\ten{8}\Msun$) requires longer
time is not obvious but can be understood from Fig.~\ref{fig:bhgrowth}b.
Consider, for instance, \BH's with masses $\sim\ten{7}\Msun$ and
$\sim\ten{10}\Msun$ at $z=0$.
Their average growth history is traced by the lower and upper
tracks in the figure.  The larger \BH\ has $\MBH\sim \ten{9}\Msun$ at $z=3$ and
has to e-fold its mass by 3 times. Conversely, the smaller \BH\ has $\MBH\sim
\ten{2}\Msun$ at $z=3$ and has to e-fold its mass by at least ten times.
The masses of large \BH's ($>\ten{8}\Msun$) increase by smaller
factors from their $z=3$ values, compared to the masses of smaller \BH's.
Thus they require shorter active phases.

In general, literature estimates of AGN lifetimes range from \ten{6} to
\ten{8}\YR\ and are still much uncertain (see the review by
\citealt{martini03}).  Models where \BH's
grow by a combination of gas accretion traced by short-lived ($\sim 10^7$\YR)
QSO activity and merging in hierarchically merging galaxies are consistent with
a wide range of observations in the redshift range $0<z<5$
(\citealt{haehnelt03}).  However it is not surprising to find a discrepancy
with our analysis since we do not consider merging and the \BH\ growth history
that we find is anti-hierarchical.  Our estimates agree better with models in
which \BH's are mainly grown by gas accretion (e.g.~\citealt{haiman03,granato04}).

With an estimate of AGN lifetime based on \BH\ demographics
similar to the one presented here \cite{yu02} find $\tau=0.3-1.3\xten{8}\YR$
for the luminous quasars using the AGN luminosity function by \cite{boyle}.
The main reason our estimate is larger is that the \cite{ueda} luminosity
function has a larger number density of objects, which results in longer
lifetimes. Indeed, the quasar lifetimes computed using the luminosity function by \cite{boyle} are $\sim 8\xten{7}\YR$ (dashed line in Fig.~\ref{fig:bhduty}b),
in agreement with \cite{yu02}.

Comparing to observational-based estimates, the
lifetimes from this paper are in agreement with the results from the length of
radio jets (see \citealt{martini03} for more details). More recently,
\cite{miller03} found from SDSS data that a very high fraction of
galaxies host an AGN ($\sim 20-40\%$) suggesting lifetimes longer than
previously thought (i.e.~$> \ten{8}\YR$).  Finally, our estimate is in
agreement with the upper limit of $\ten{9}\YR$ set by the timescale over
which the quasar luminosity density rises and falls (see, e.g., \citealt{osmer03}).

In summary, we estimate that local high mass BH's ($M>\ten{9}\Msun$) have been
active, on average, $\sim 1.5\xten{8}\YR$.  On the contrary, the assembly of
lower mass \BH's has required active phases lasting at least three times
that much
($\sim 4.5\xten{8}\YR$).  These average lifetimes can be as large as
$\sim \ten{9}$ if one
considers the smaller efficiency and fraction of Eddington luminosity  which
are still compatible with local \BH's ($\eff=0.04$, $\lambda=0.1$).

\section{Conclusions}
\label{sec:conclusions}

We have quantified the importance of mass accretion during AGN phases in the
growth of supermassive black holes (\BH) by comparing the mass function of
black holes in the local universe with that expected from AGN relics, which are
black holes grown entirely during AGN phases.

The local \BH\ mass function (\BHMF) has been estimated by applying the
well-known correlations between \BH\ mass, bulge luminosity and stellar
velocity dispersion to galaxy luminosity and velocity functions.  We have found
that different \BH-galaxy correlations provide the same \BHMF\ only if they
have the same intrinsic dispersion, confirming the findings of \cite{MH03}.
The
density of supermassive black holes in the local universe is $\rhoBH =
4.6_{-1.4}^{+1.9}\, h_{0.7}^2 \rhoBHunits$.

The relic \BHMF\ is derived from the continuity equation with the only assumption
that AGN activity is due to accretion onto massive \BH's and that merging is
not important.  We find that the relic \BHMF\ at $z=0$ is generated mainly at
$z<3$ where the major part of \BH's growth takes place.  The relic \BHMF\ at
$z=0$ is very little dependent on its value at $\zo=3$ since the main growth of
\BH's took place at $z<3$. Moreover, the \BH\ growth is anti-hierarchical in
the sense that smaller \BH's ($M_{BH}<10^{8}M_\odot$) grow at lower redshifts
($z<1$) with respect to more massive one's ($z\sim 1-3$). If the correlations
between \BH\ mass and host-galaxy-properties hold at higher redshifts this 
would represent a potential problem
for hierarchical models of galaxy formation.

Unlike previous work, we find that the \BHMF\ of AGN relics is perfectly
consistent with the local \BHMF\ indicating the local black holes were mainly
grown during AGN activity.  This agreement is obtained while satisfying, at the
same time, the constraints imposed from the X-ray background both in terms of
BH mass density and fraction of obscured AGN's. 
The reasons for the solution of the discrepancy at high masses found by 
other authors are  the following:
\begin{itemize}
\item we have taken into account the intrinsic dispersion of the
\MBHsig\ and \MBHlum\ correlations in the determination of the local \BHMF;
\item we have adopted the coefficients of the \MBHsig\ and \MBHlum\
($K$ band) correlations
derived by \cite{MH03} after considering only `secure' BH masses;
\item we have derived improved bolometric corrections which do not take into
account reprocessed IR emission in the estimate of the bolometric luminosity. 
\end{itemize}

The comparison between the local and relic \BHMF's also suggests that the
merging process at low redshifts ($z<3$) is not important in shaping the relic
\BHMF, and allows us
to estimate the average radiative efficiency ($\varepsilon$),
the ratio between emitted and Eddington luminosity ($\lambda$) and the average
lifetime of active \BH's.

Our analysis thus suggests the following scenario: local black holes grew
during AGN phases in which accreting matter was converted into radiation with
efficiencies $\varepsilon = 0.04-0.16$ and emitted at a fraction $\lambda =
0.1-1.7$ of the Eddington luminosity. The average total lifetime of these
active phases ranges from $\simeq 4.5\times 10^{8}$ yr for
$M_{BH}<10^{8}M_\odot$ to $\simeq 1.5\xten{8}$ yr for
$M_{BH}>10^{9}M_\odot$ but can become as large as $\sim \ten{9}\YR$
for the lowest acceptable $\eff$ and $\lambda$ values.
\section*{Acknowledgments}
We thank Reinhard Genzel, Linda Tacconi and Andrew Baker for useful suggestions.
A.M. and L.K.H. acknowledge support by MIUR (Cofin01-02-02).
A.M., G.R., and R.M., acknowledge support by INAOE, Mexico, during the
2003 Guillermo-Haro Workshop where part of this work was performed.
This research has made use of NASA's Astrophysics Data System Bibliographic Services.


\label{lastpage}
\end{document}